\documentclass[sigconf,authorversion,nonacm]{acmart}

\usepackage[utf8]{inputenc}

\usepackage{cprotect}
\usepackage{graphicx}
\usepackage{dcolumn}
\usepackage{bm}
\usepackage{amsmath}
\usepackage{amsfonts}
\usepackage{listings}
\usepackage{xcolor, colortbl}
\usepackage{algorithm2e}
\usepackage[textsize=small]{todonotes}
\usepackage{float}
\usepackage{tabularx}
\usepackage{enumerate}
\usepackage{chemformula}

\definecolor{codegreen}{rgb}{0,0.3,0}
\definecolor{codered}{rgb}{0.3,0,0}
\definecolor{codeblue}{rgb}{0,0,0.3}
\definecolor{codegray}{rgb}{0.3,0.3,0.3}
\definecolor{codepurple}{rgb}{0.58,0,0.82}
\definecolor{backcolour}{rgb}{0.95,0.95,0.92}

\definecolor{codegreen}{rgb}{0,0.6,0}
\definecolor{codegray}{rgb}{0.5,0.5,0.5}
\definecolor{codepurple}{rgb}{0.58,0,0.82}
\definecolor{backcolour}{rgb}{0.95,0.95,0.92}
\definecolor{lightred}{rgb}{0.97, 0.81, 0.8}
\definecolor{lightgreen}{rgb}{0.84, 0.91, 0.83}
\definecolor{lightyellow}{rgb}{1,0.95,0.8}
\definecolor{lightblue}{rgb}{0.85, 0.91, 0.99}
\definecolor{lightgray}{rgb}{0.95,0.95,0.95}

\lstdefinestyle{CodeStyle}{
    backgroundcolor=\color{backcolour},
    commentstyle=\color{codegreen},
    keywordstyle=\color{magenta},
    numberstyle=\tiny\color{codegray},
    stringstyle=\color{codepurple},
    basicstyle=\ttfamily\footnotesize,
    breakatwhitespace=false,
    breaklines=true,
    keepspaces=true,
    numbers=left,
    numbersep=8pt,
    showspaces=false,
    showstringspaces=false,
    showtabs=false,
    tabsize=2,
    frame=single,
    framerule=1pt,
    resetmargins=true,
    xleftmargin=2pt,
    xrightmargin=2pt,
    basicstyle=\fontsize{7}{7}\selectfont\ttfamily
}
\begin{document}

\title{Hybrid quantum programming with \\PennyLane Lightning on HPC platforms}

\author{Ali Asadi}
\affiliation{
  \institution{Xanadu Quantum Technologies Inc.}
  \streetaddress{777 Bay St.}
  \city{Toronto}
  \state{Ontario}
  \country{Canada.}
}

\author{Amintor Dusko}
\affiliation{
  \institution{Xanadu Quantum Technologies Inc.}
  \streetaddress{777 Bay St.}
  \city{Toronto}
  \state{Ontario}
  \country{Canada.}
}

\author{Vincent Michaud-Rioux}
\affiliation{
  \institution{Xanadu Quantum Technologies Inc.}
  \streetaddress{777 Bay St.}
  \city{Toronto}
  \state{Ontario}
  \country{Canada.}
}

\author{Chae-Yeun Park}
\affiliation{
  \institution{Xanadu Quantum Technologies Inc.}
  \streetaddress{777 Bay St.}
  \city{Toronto}
  \state{Ontario}
  \country{Canada.}
}

\author{Isidor Schoch}
\affiliation{
  \institution{Xanadu Quantum Technologies Inc.}
  \streetaddress{777 Bay St.}
  \city{Toronto}
  \state{Ontario}
  \country{Canada.}
}
\additionalaffiliation{
  \institution{ETH Zurich}
  \city{Zurich}
  \country{Switzerland.}
}
\authornote{Work undertaken while at Xanadu.}

\author{Shuli Shu}
\affiliation{
  \institution{Xanadu Quantum Technologies Inc.}
  \streetaddress{777 Bay St.}
  \city{Toronto}
  \state{Ontario}
  \country{Canada.}
}

\author{Trevor Vincent}
\affiliation{
  \institution{Xanadu Quantum Technologies Inc.}
  \streetaddress{777 Bay St.}
  \city{Toronto}
  \state{Ontario}
  \country{Canada.}
}
\authornote{Work undertaken while at Xanadu.}

\author{Lee J. O'Riordan}
\affiliation{
  \institution{Xanadu Quantum Technologies Inc.}
  \streetaddress{777 Bay St.}
  \city{Toronto}
  \country{Canada.}
}
\email{lee@xanadu.ai}
\authornote{Corresponding author.}

\renewcommand{\shortauthors}{A. Asadi, A. Dusko, C. Y. Park, V. Michaud-Rioux, I. Schoch, S. Shu, T. Vincent, L. J. O'Riordan}

\date{March, 2024}

\begin{abstract}

We introduce PennyLane's Lightning suite, a collection of high-performance state-vector simulators targeting CPU, GPU, and HPC-native architectures and workloads. Quantum applications such as QAOA, VQE, and synthetic workloads are implemented to demonstrate the supported classical computing architectures and showcase the scale of problems that can be simulated using our tooling. We benchmark the performance of Lightning with backends supporting CPUs, as well as NVidia and AMD GPUs, and compare the results to other commonly used high-performance simulator packages, demonstrating where Lightning's implementations give performance leads. We show improved CPU performance by employing explicit SIMD intrinsics and multi-threading, batched task-based execution across multiple GPUs, and distributed forward and gradient-based quantum circuit executions across multiple nodes. Our data shows we can comfortably simulate a variety of circuits, giving examples with up to 30 qubits on a single device or node, and up to 41 qubits using multiple nodes.

\end{abstract}
\settopmatter{printfolios=true}
\maketitle

\section{Introduction}\label{sec:intro}

The PennyLane quantum programming library~\cite{bergholm2022pennylane} offers a device (CPU, GPU, QPU, xPU) agnostic approach to quantum computing, wherein physical or simulated devices can be programmed from the same interface. As Python is rarely the default choice for performant scientific workloads, PennyLane offers a collection of C++-backed simulators to complement the ecosystem, written using modern C++ (17/20). As the goal is to enable better workload performance, whilst also supporting a wide range of classical hardware ecosystems, we build support for multiple hardware backends from a common framework. PennyLane's focus, while supporting all currently accessible quantum hardware and software platforms, has always been to ensure support for differentiable programming of quantum circuits, and tight integration with industry and research machine-learning frameworks, such as JAX, PyTorch, TensorFlow, and Autograd~\cite{maclaurin2015autograd, paszke2017automatic, tensorflow2015, jax2018github}. 

Batch-transformed quantum circuit methods, examples of which include parameter-shifted gradients~\cite{Mitarai_2018, Schuld_2019, Wierichs_2022}, circuit-cutting~\cite{Lowe2023fastquantumcircuit}, and the metric tensor transform~\cite{meyer2021}, often require multiple executions of circuits to evaluate the quantities of interest, so ensuring fast end-to-end runs of multiple (and often concurrent) workloads is paramount. As not all workloads are equal, having support for a variety of different paradigms with optimizations for each, can allow the best overall time-to-solution. Considering a set of common workloads such as the following:
\begin{itemize}
   \item Task-based executions of batched and independent circuits running on limited resources per circuit.
   \item Circuits with full access to CPU sockets or local GPUs per execution.
   \item Large circuits that cannot fit on a single resource, such as with multi-GPU and/or multi-node workloads.
   \item Combinations of the above in a way that efficiently uses all resources to solve large complex problems.
\end{itemize}

We will aim to address and demonstrate solving the first three of the above workloads using PennyLane's Lightning devices. For completeness, we also demonstrate how a user of a multi-device quantum system (hardware or simulator, local or remote), can define workloads to be subdivided and solved concurrently, and do so efficiently for each respective component of the problem.

This paper is structured as follows: Section~\ref{sec_pl_arch} begins with a discussion of the Lightning simulator suite's respective architecture, including gate implementations, designs and features; Section~\ref{sec_performance} will focus on comparative runtimes and demonstrating workloads across the gamut from single-qubit gates to hybrid classical-quantum workloads; Section~\ref{sec_heterogeneous} will discuss the use of PennyLane in large-scale computing environments, with a focus on HPC systems; Section~\ref{sec_summary} will conclude with a discussion on future strategies and execution environments for quantum workloads coupled to HPC systems.

\section{PennyLane Lightning architecture}\label{sec_pl_arch}

Several of the foreseen applications of quantum computing are, in essence, optimization problems where the cost function is obtained by evaluating a quantum circuit with parametrized gates. It is crucial for any such algorithm to be able to compute the gradient of the cost function as efficiently as possible.
PennyLane's default simulator device {\emph{default.qubit}} is an end-to-end Python implementation that ties into automatic differentiation (AD) frameworks such as PyTorch, TensorFlow, and JAX~\cite{jax2018github, paszke2017automatic, tensorflow2015}.
This is a very powerful paradigm wherein we can natively offload any part of the stack to the respective framework, such as JIT compilation with JAX or native GPU offload with PyTorch.
However, gate and measurement AD comes with a large memory overhead as the qubit count and circuit depth increases.
For example, simulating a circuit with 24 qubits and thousands of parameters can already require hundreds of gigabytes of memory.

To overcome this, PennyLane also provides interfaces to these AD frameworks, allowing a device to define its own custom gradient rules; the AD frameworks then treat each device execution as a black-box, where parameters go in, and results come out, with the device specifying how it handles gradient-based execution. For this, the native operations run by each respective device can be abstracted away from the framework's native traceable calls, and one can simply define a custom vector-Jacobian product (VJP), or Jacobian-vector product (JVP), depending on whether forward-mode or reverse-mode differentiation is required. 

The above abstraction allows us to define a device in any means we wish (hardware or simulator), and provide custom gradient rules for said device.
Considering simulators, PennyLane offers a suite of high-performance simulators written in C++ and targeting a variety of CPU and GPU platforms named ``PennyLane Lightning''~\cite{pennylane_lightning_repo}.
The suite is comprised of device module backends written using C++20. As such, the Lightning suite can be compiled to run on any available system \& target device, assuming a supporting compiler and third-party libraries exist (e.g., GCC 11+, Clang 14+, MSVC 2022, etc).

The suite defines higher-level API components to allow support in Python through PennyLane, or externally as a series of C++ state vector simulators. The suite is currently comprised of the following backends (see Figure~\ref{fig:lightning_arch}):

\begin{figure*}
\includegraphics[width=0.75\linewidth]{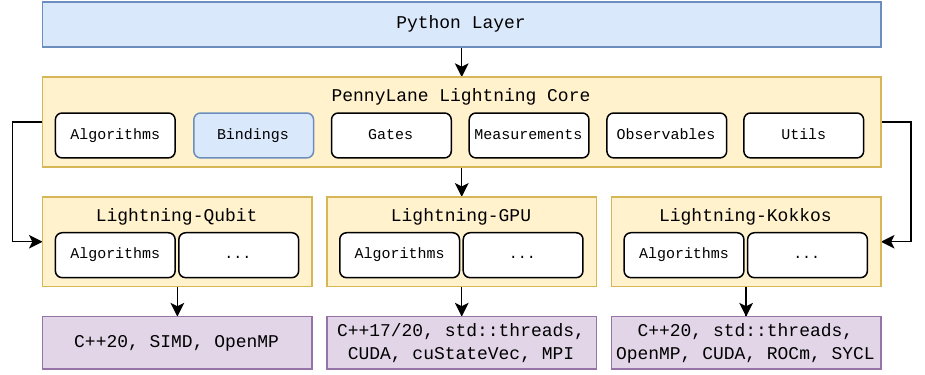}
\caption{PennyLane Lightning template module architecture. Each Lightning backend device follows the same architectural design, with differences in the implemented gate set, observables, and compile-time targets. The common modules allow us to easily implement functionality across the package ecosystem and allow specialization per backend target.}
\label{fig:lightning_arch}
\end{figure*}

\begin{itemize}
  \item Lightning-Qubit: This is the widely-supported C++-backed device, and runs on Windows, MacOS and Linux across x86\_64, ARM64/AArch64 and PPC64LE (depending on support by the given operating system), and will be installed alongside the PennyLane Python library. This can be accessed as \verb+lightning.qubit+ through PennyLane's device plugin interface.
  \item Lightning-GPU: This device builds upon the architecture of \verb+lightning.qubit+, but allows direct offloading to the NVIDIA cuQuantum SDK~\cite{cuquantum} for state vector simulation, working on CUDA SM~7.0~\cite{cuda} and newer hardware devices. This can be accessed as \verb+lightning.gpu+ through PennyLane's device plugin interface.
  \item Lightning-Kokkos: This device features a direct port of Lightning-Qubit's kernel algorithms to Kokkos~\cite{kokkos-1, kokkos-2}, and allows us to directly target OpenMP or \verb+std::threads+ accelerated gate targets on CPU devices, as well as execution on both AMD and NVIDIA GPUs directly through ROCm~\cite{rocm57} and CUDA SDKs respectively. This can can be accessed as \verb+lightning.kokkos+ through PennyLane's device plugin interface.
\end{itemize}

The supported backends and platforms are summarized in Tab.~\ref{tab:avail}. All Lightning devices are fully templated end-to-end, allowing support for either 32-bit or 64-bit floating point operations, both at the Python layer and explicitly at the C++ layer. Compiled binaries are distributed on the Python Packaging Index (PyPI, our official distribution platform) and Conda-Forge (CF). 
Docker images are available on DockerHub~\cite{dockerhub_url}, which can be useful to install the simulators on HPC systems.
Certain backends and options still require to be built from source, which is indicated by \texttt{src}. Any platform with a binary distribution is also implicitly supported by the built-from-source option, and those listed are currently the ones we have validated to build and run successfully.
Yet another option to install the Lightning plugins is \texttt{spack}~\cite{spack}, a package manager that makes it convenient to configure, build and install HPC software.

\begin{table*}[h]
\begin{tabularx}{0.8\textwidth}{
|>{\centering\arraybackslash}X
|>{\centering\arraybackslash}X
|>{\centering\arraybackslash}X
|>{\centering\arraybackslash}X
|>{\centering\arraybackslash}X
|>{\centering\arraybackslash}X
|>{\centering\arraybackslash}X | }
\hline
\hline

          & L-Qubit                          &  L-GPU                           & L-GPU (MPI)                          & L-Kokkos (OMP)                       & L-Kokkos (CUDA)                      & L-Kokkos (HIP)           \\
\hline
Linux x86 & \cellcolor{lightgreen}PyPI/CF/DH & \cellcolor{lightgreen}PyPI/CF/DH & \cellcolor{lightyellow} \texttt{src} & \cellcolor{lightgreen}PyPI/CF/DH     & \cellcolor{lightgreen}DH             & \cellcolor{lightgreen}DH \\
\hline
Linux ARM & \cellcolor{lightgreen}PyPI/CF    & \cellcolor{lightgreen}CF         & \cellcolor{lightyellow} \texttt{src} & \cellcolor{lightgreen}PyPI/CF        & \cellcolor{lightyellow} \texttt{src} & \cellcolor{lightgray}    \\
\hline
Linux PPC & \cellcolor{lightgreen}PyPI/CF    & \cellcolor{lightgreen}CF         & \cellcolor{lightyellow} \texttt{src} & \cellcolor{lightgreen}PyPI/CF        & \cellcolor{lightyellow} \texttt{src} & \cellcolor{lightgray}    \\
\hline
MacOS x86 & \cellcolor{lightgreen}PyPI/CF    &\cellcolor{lightgray}             &  \cellcolor{lightgray}               & \cellcolor{lightgreen}PyPI/CF        & \cellcolor{lightgray}                & \cellcolor{lightgray}    \\
\hline
MacOS ARM & \cellcolor{lightgreen}PyPI/CF    &\cellcolor{lightgray}             & \cellcolor{lightgray}                & \cellcolor{lightgreen}PyPI/CF        & \cellcolor{lightgray}                & \cellcolor{lightgray}    \\
\hline
Windows   & \cellcolor{lightgreen}PyPI       &\cellcolor{lightgray}             &  \cellcolor{lightgray}               & \cellcolor{lightyellow} \texttt{src} & \cellcolor{lightgray}                & \cellcolor{lightgray}    \\
\hline
\hline
\end{tabularx}
\caption{PennyLane Lightning is distributed on the Python Packaging Index (PyPI) and Conda-Forge (CF). Linux x86 Docker images are also available on DockerHub (DH). Certain versions like the distributed Lightning-GPU backend must be built from source (src).}
\label{tab:avail}
\end{table*}

\cprotect\subsection{\verb+lightning.qubit+}

\verb+lightning.qubit+ is PennyLane's ``supported everywhere'' C++ device, which serves as the default performance-focused simulator provided for PennyLane. The device follows the PennyLane plugin architecture, providing a compiled Pybind11~\cite{pybind11} module to be imported for use with a Pythonic frontend, or directly as a CMake-enabled C++ simulator.
The core elements of the simulator focus strictly on standards-compliant C++17/20 code, with optional OpenMP~\cite{chandra2001parallel} additions to improve gradient evaluations over multiple circuit observables using the adjoint gradient method~\cite{jones2020efficient}, with parallelized gate-level kernels as an additional compile-in supported option. This device features intrinsic-backed computations on x86\_64 platforms, with a dispatch model allowing automated calling to default (LM), AVX-2 or AVX-512~\cite{jeffers2016intel} kernels without need for recompilation, depending on the queried support at runtime from the CPU. 

\subsubsection{Implementing kernels for quantum gates}

\verb+lightning.qubit+'s native focus is to support PennyLane workloads (including quantum circuit gradients) and ensuring that each requested gate application operates as efficiently as possible.
Lightning's default gate kernels (currently supported across x86\_64, ARM, PPC64LE) are essentially contracting a relatively small sparse or dense matrix (the gate) with a multidimensional tensor (the state vector).
This is efficiently done using bitwise operations to recover the state vector coefficients on the fly and implementing the gate without constructing intermediate vectors or matrices. To ensure generality, we also have generic single, two, and many-qubit gates as fall-backs for when arbitrary matrix-based gate applications are required.

\RestyleAlgo{ruled}
\begin{algorithm}
\KwData{$\textrm{SV}$, $0\leq q < \log_2{|SV|}$, $f( SV, i_0 \in \mathbb{N}_0, i_1 \in \mathbb{N}_0 )$}
$\textrm{n\_qubits} \gets \log_2{|SV|}$\;
$\textrm{bit\_width} \gets 64$\; 
$\textrm{q\_offset} \gets \textrm{n\_qubits} - q - 1$\; 
$\textrm{stride} \gets 1 \ll \textrm{q\_offset}$\;
$\textrm{mask\_high} \gets (\textrm{uint\_max} \ll (\textrm{q\_offset} + 1))~\textbf{\&}~\textrm{uint\_max}$\; 
$\textrm{mask\_low} \gets \textrm{uint\_max} \gg (\textrm{bit\_width} - \textrm{q\_offset})$\;
\For{$k\gets0$ \KwTo $2^{(\textrm{\normalfont n\_qubits}-1)}$}{
    $i_0 = ((2\times k)~\textbf{\&}~\textrm{mask\_high}) ~\textbf{|}~ (\textrm{mask\_low} ~\textbf{\&}~ k)$\;
    $i_1 = i_0 ~\textbf{|}~ \textrm{stride}$\;
    $f$(SV, $i_0$, $i_1$);
}
\caption{Generic single-qubit gate}\label{alg:generic_1q_gate}
\end{algorithm}

A generic algorithm for single-qubit gates is defined by Alg.~\ref{alg:generic_1q_gate}, which follows a design common to many such state vector methods~\cite{qulacs_2021, vanbeeumen2023qclab, qibojit_paper, quest_2019}.
We start by identifying the offsets and strides to access the interacting state vector coefficients for a given gate application, apply the update operation on the coefficients, and repeat the process on the next pair.
Interacting pairs of coefficients in the for-loop are strictly disjoint, making it possible to evaluate these updates concurrently using multi-threaded execution, which remains as an opt-in compile-time option for \verb+lightning.qubit+.
The difference between each gate lies in the so-called coefficient interaction function, $f$. The coefficient interaction functions for Pauli-X and the Phase gate are given by Listing~\ref{lst:func_paulix_phase}. Both functions are templated by \verb+T+ to allow use of both 32-bit (single) and 64-bit (double) precision representations.

\begin{minipage}{\linewidth}

\begin{lstlisting}[caption={Pauli-X (top) and Phase-shift (bottom) gate coefficient interaction functions.}, label={lst:func_paulix_phase}, style=CodeStyle,language=C++, captionpos=b]
auto PauliX = [](std::complex<T> *SV, 
                 size_t i0, size_t i1) {
    std::swap(SV[i0], SV[i1]);
};
...
const std::complex<T> s = 
  inverse ? 
  std::exp(-std::complex<T>{0, angle}) : 
  std::exp( std::complex<T>{0, angle});
auto PShift = [&s](std::complex<T> *arr, 
                   size_t i0, size_t i1) { 
    arr[i1] *= s; 
};
\end{lstlisting}
\end{minipage}

Another crucial algorithm is the controlled single-qubit gate algorithm defined by Alg.~\ref{alg:controlled_1q_gate}.

\RestyleAlgo{ruled}
\begin{algorithm}
\KwData{$\textrm{SV}$, $0\leq q < \log_2{|SV|}$, ctrls $\in \{0, \log_2{|SV|} - 1\}$, $f( SV, i_0 \in \mathbb{N}_0, i_1 \in \mathbb{N}_0 )$}
$\textrm{n\_qubits} \gets \log_2{|SV|}$\;
$\textrm{n\_ctrls} \gets |\textrm{ctrls}|$\;
$\textrm{bit\_width} \gets 64$\; 
$\textrm{q\_offset} \gets \textrm{n\_qubits} - q - 1$\; 
$\textrm{stride} \gets 1 \ll \textrm{q\_offset}$\;
$\textrm{ctrl\_offsets} \gets \textrm{n\_qubits} - $ctrls$ - 1$\;
$\textrm{ctrl\_strides} \gets 1 \ll \textrm{ctrl\_offsets}$\;
$\textrm{masks} \gets $getMasks$(\textrm{ctrl\_offsets}, \textrm{q\_offset})$\;
\For{$k\gets 0$ \KwTo $2^{(\textrm{\normalfont n\_qubits}-1-\textrm{\normalfont n\_ctrls})}$}{
    $i_0 = k ~\textbf{|}~ \textrm{masks}[0]$\;
    \For{$i\gets 1$ \KwTo $\textrm{\normalfont n\_ctrls}$}{
	    $i_0 = i_0 ~\textbf{|}~ ((k \ll i) ~\textbf{\&}~ \textrm{masks}[i])$\;
    }
    \For{$i\gets 0$ \KwTo $\textrm{\normalfont n\_ctrls} - 1$}{
	    $i_0 = i_0 ~\textbf{|}~ \textrm{ctrl\_strides}[i]$\;
    }
    $i_1 = i_0 ~\textbf{|}~ \textrm{stride}$\;
    $f$(SV, $i_0$, $i_1$);
}
\caption{Generic controlled single-qubit gate}\label{alg:controlled_1q_gate}
\end{algorithm}

In Alg.~\ref{alg:controlled_1q_gate}, \verb+getMasks+ is a function that concatenates and sorts the input into \verb+bits+, and then returns $n + 1$ bit masks with:
\begin{enumerate}
  \item \verb+masks[0]+ having trailing ones up to \verb+bits[0]+ (not included);
  \item \verb+masks[i]+ having ones between \verb+bits[i-1]++\verb+1+ (included) and \verb+bits[i]+ (not included);
  \item \verb+masks[n]+ having leading ones up to \verb+bits[n]+ (included).
\end{enumerate}
An $n$--controlled single-qubit gate falls under the umbrella of $n+1$--qubit unitary gates.
While it is possible to apply controlled gates using the generic gate algorithm, this is wasteful because we only need to consider state vector entries where all control bits are equal to 1 (or a prescribed bit string).
Instead of computing $2^{n+1}$ state vector indices and performing a $2^{n+1}\times 2^{n+1}$ matrix product onto the corresponding coefficients, only 2 indices and coefficients are required for every $k$ in the loop.
One can see from the upper bound that the loop is reduced by a factor $2^{-n}$ compared with the non-controlled single-qubit gate kernel, and hence we expect the time to apply an $n$--controlled gate to be rougly $2^{-n}$ the time to apply the non-controlled version.

Alg.~\ref{alg:generic_1q_gate} and Alg.~\ref{alg:controlled_1q_gate} can be generalized to an arbitrary number of qubits. The Lightning backends implement specialized kernels for 1- up to 4-qubit gates, beyond which a completely general implementation takes over.

\subsubsection{SIMD accelerated gates}

As an extension of the above implementation, one can also consider scenarios where operations can be SIMD vectorised -- in this case, the use of intrinsic operations can help to improve the intermediate range accesses: when strides are wide enough to allow multiple loads and updates in SIMD registers, we can improve performance by calling explicitly targeted gate kernels. For every $k$ floating-point values, we can represent $k/2$ complex numbers. Depending on the access and interactions required, given the strides between interacting values, we consider multiple function implementations to realise these gates:
\begin{enumerate}
  \item Interacting coefficients are on the same register: internal interactions only (intra-register/lane).
  \item Interacting coefficients are on different registers: external interactions only (cross-register/lane).
  \item For multi-qubit gates, interactions are a mix of internal and external interactions (both intra and cross-lane).
\end{enumerate}

For both AVX2 (256-bit registers) and AVX-512 (512-bit registers), and given that we require either 64-bits/128-bits respectively to represent each complex number, we are limited to storing either 8/4 values for AVX2 or 16/8 for AVX-512 (assuming single/double precision).

The complexity of the implementation grows with the number of qubits required for a given gate. With this lies the challenge of exhaustively implementing SIMD kernels - we will always require both intra-lane and inter-lane operations to realise the gates, and the number of implementations grows with the number of interacting qubits. For a single qubit gate, inter and intra operations alone can implement the gate-set.
For two-qubit gates, we require separate operations per qubit index being interacted with, requiring up to four separate functions: inter-inter, intra-intra, inter-intra, and intra-inter (for symmetric two-qubit gates, both inter-intra and intra-inter are equivalent, and we can reduce the count by 1). For a general implementation, such as for an arbitrary $n$-qubit gate we require $2^n$ functions at the worst case to realise the full implementation.

\begin{figure}
  \centering
  \includegraphics[width=\linewidth, trim={0.25cm 2.5cm 0 2cm},clip]{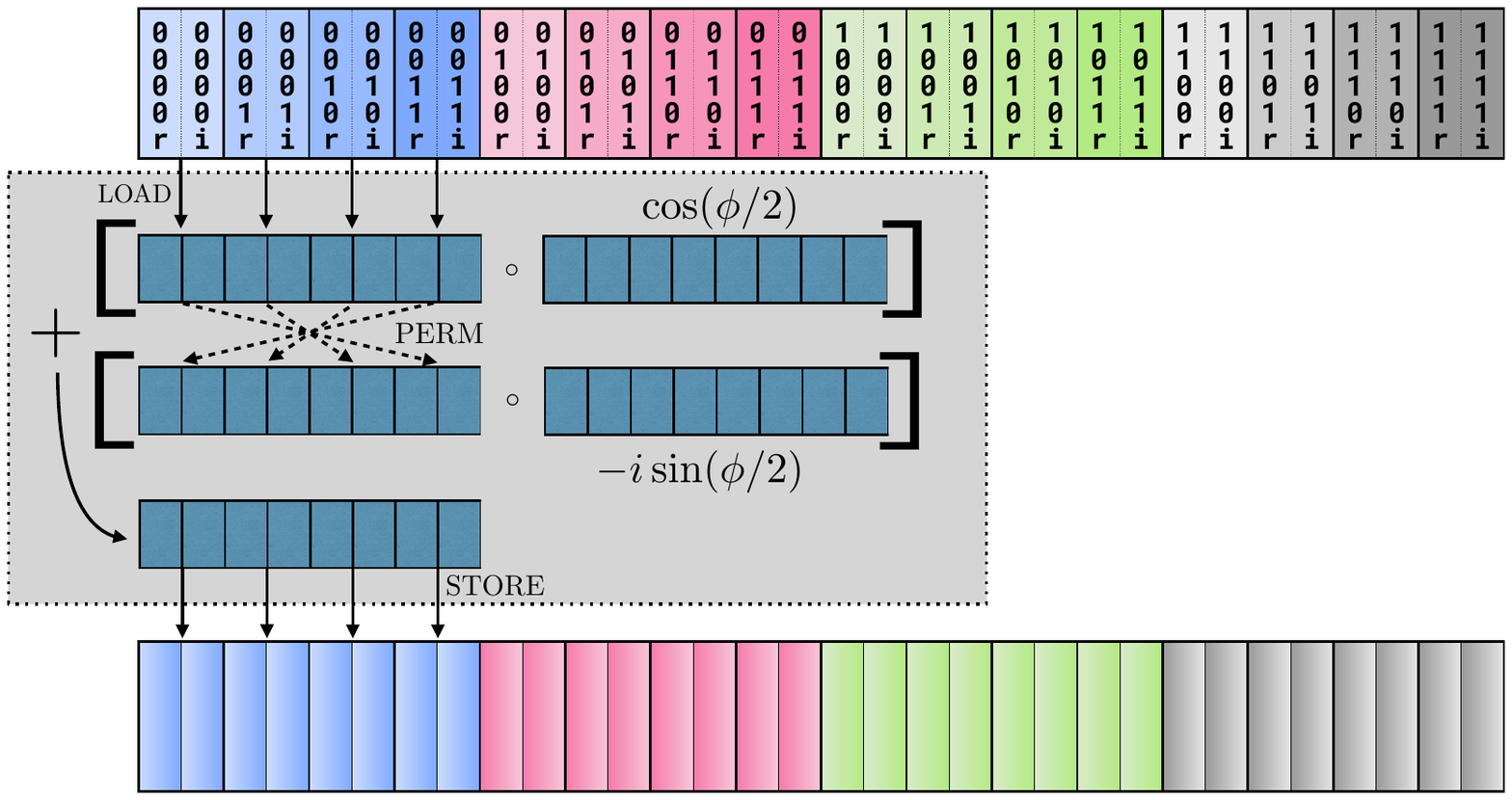}
  \caption{AVX-512 kernel stages for applying an IsingXX gate across the intra-intra (coefficients all reside within a single register) qubit indices. Data is loaded into the registers, a permuted copy is created in a separate register, and the required trigonometric values are elementwise multiplied to produce the desired gate kernel outputs, which are then summed together. Upon completion, the register data overwrites the associated indices in main memory, and the process repeats on the next set of coefficients.}
  \label{fig:isingxx_application_AVX512}
\end{figure}

While possible to generalise to arbitrary gates, the added complexity may not be worth the cost unless a given workload requires a well-optimized gate structure. For larger multi-qubit kernels, it may be more beneficial to offload to BLAS functions, and we expect this will be true for 4/5 qubits and beyond. To date, we have implemented intrinsic kernels for the well-known single qubit constant (X, Y, Z, H, S, T) and parametric (Phase, RX, RY, RZ, Rot) gates, their controlled (two-qubit) counterparts, as well as some additional specialised two-qubit parametric gates (SWAP, IsingXX, IsingXY, IsingYY, IsingZZ). Additionally, we have implemented the respective generators of the above parametric gates to be used for gradient evaluations. Figure~\ref{fig:isingxx_application_AVX512} demonstrates the process of applying an IsingXX gate, defined as
\begin{eqnarray}
\label{eqn:isingxx}
\begin{split}
XX(\phi) &= \exp\left(-i \frac{\phi}{2} (X \otimes X)\right) \\
  & = \begin{bmatrix}
      \cos(\phi / 2) & 0 & 0 & -i \sin(\phi / 2) \\
      0 & \cos(\phi / 2) & -i \sin(\phi / 2) & 0 \\
      0 & -i \sin(\phi / 2) & \cos(\phi / 2) & 0 \\
      -i \sin(\phi / 2) & 0 & 0 & \cos(\phi / 2)
\end{bmatrix},\end{split}
\end{eqnarray}
on qubits 0 and 1 for the AVX-512 implementation, which shows full intra-register use for the given indices. While not shown, the AVX2 counterpart implementation at the same precision crosses the inter-register boundary for the interacting coefficients of this gate, and will need to make use of cross-lane permutations. Depending on the given precision (single, double), and choice of indices, the respective intra register and inter (lane crossing) functions will be called to support the given operations to realise the gate application.

\cprotect\subsection{\verb+lightning.gpu+}

Often directly targeting GPUs requires the definition of either custom kernels representing the operations, or appropriate wrapping around BLAS-like function calls to perform matrix-vector updates on a state vector with the numerical gate representations. As with \verb+lightning.qubit+, we opted for a C++17 (and more recently, C++20) implementation that provides all of the necessary framework to manage GPU memory, mediate CPU-GPU and GPU-GPU transfers, and support dynamic GPU device selection. For this, \verb+lightning.gpu+ was created as a fork of \verb+lightning.qubit+, and ties directly into the NVIDIA cuQuantum SDK cuStateVec library~\cite{cuquantum} for state vector coefficient updates and measurement process evaluations, including expectation values, bit-string samples, probabilities, and variances.

The adjoint differentiation gradient method was implemented to work directly with GPU-defined state vectors to ensure the device works with PennyLane's gradient-first approach. This method is directly compatible with all C++-defined parametric gates, and with observables such as single gates, Pauli words, Hamiltonians~(\verb+qml.Hamiltonian+), and even general sparse matrices in PennyLane~(\verb+qml.SparseHamiltonian+).
This device can also operate as a C++ frontend for the NVIDIA cuQuantum cuStateVec library, offering the same feature set as the device's Python interface.

To avoid unnecessary transfers between host and device when applying matrices for unsupported gates in the cuStateVec API (as of the time of writing such as the non-parametric Pauli gates), a caching mechanism is employed allowing constant gates to be stored on first application. These are then reused from a device-side buffer thereafter, avoiding the need for additional CPU-GPU transfers. 

\begin{minipage}{\linewidth}
\begin{lstlisting}[label={lst:lgpu_ham}, style=CodeStyle,language=Python, caption={Multi-GPU Hamiltonian expectation value gradient evaluation. The gradient of the provided cost function with inner function, $f$, will offload to the multi-GPU batched adjoint differentiation pipeline.}, captionpos=b]
n_qubit = 20
batch_obs = True # int values control the batch-size
dev = qml.device("lightning.gpu", wires=n_qubit, batch_obs=batch_obs) 

terms = [qml.PauliX(i%n_qubit) @ qml.PauliY((i+1)%n_qubit) for i in range(n_qubit)]
coeffs = np.random.rand(n_qubit)
H = qml.Hamiltonian(coeffs = coeffs, terms = terms)
...
@qml.qnode(dev, diff_method="adjoint")
def circuit(params):
    ...
    return qml.expval(H)
...
def cost_fn(params):
    return f(circuit(params))

g = qml.grad(cost_fn)(params)
\end{lstlisting}
\end{minipage}

The device natively supports batched execution of observables across multiple GPUs on a single node (useful for many-term Hamiltonian simulations).
A concurrent execution pipeline was added to allow the distribution of multiple observables over locally available GPU devices with the addition of a single-producer-multi-consumer task queue. For a host system having $g$ GPUs and aiming to evaluate a Hamiltonian expectation value, $\langle \mathcal{H} \rangle$, where $\mathcal{H} = \sum_{i=1}^{n} H_i$; each item of work can be offloaded to an accessible GPU evenly in chunks of $n/g$, or as a user-defined batch-size $b$ of $b/g$, which are reduced over to produce the final expectation value. This allows for a pipeline to be tuned to best suit the system size over the required number of terms, with preallocation for all required memory (1 state vector copy per observable), or in user-defined sizes to lower the memory footprint, at the expense of additional compute overhead during the adjoint gradient evaluations. This can be enabled with the example of Listing~\ref{lst:lgpu_ham}.

\begin{minipage}{\linewidth}
\begin{lstlisting}[label={lst:lgpu_mpi}, style=CodeStyle, language=Python, caption={MPI-distributed circuit example using 36 qubits returning probabilities to the root process.}, captionpos=b]
from mpi4py import MPI
import pennylane as qml
import numpy as np

comm = MPI.COMM_WORLD
rank = comm.Get_rank()
num_qubits = 36
dev = qml.device('lightning.gpu', wires=num_qubits, mpi=True)
prob_wires = [0, num_qubits//2, num_qubits-1]

@qml.qnode(dev)
def mpi_circuit():
    qml.Hadamard(0)
    for i in range(1, num_qubits):
        qml.CNOT(wires=[i-1, i])
    return qml.probs(wires=prob_wires)

local_probs = mpi_circuit()
recv_counts = comm.gather(len(local_probs), root=0)
probs = np.zeros(2**len(prob_wires)) if rank == 0 else None
comm.Gatherv(local_probs, [probs, recv_counts], root=0)
\end{lstlisting}
\vspace{-0.2em}
\end{minipage}

In addition to single node workloads, \verb+lightning.gpu+ allows for fully distributed simulation of workloads through PennyLane via MPI, Lightning-GPU's C++ module bindings and the NVIDIA cuStateVec distributed API functions. Support currently exists for probabilities, sample-based workloads, and the various expectation values as supported by the single-node design, including support for a distributed CSR sparse-matrix representation of the observable. Harnessing mpi4py for its convenient user-facing layer to both initialise the MPI runtime and query ranks, the end user is free to build workloads as reported in Listing~\ref{lst:lgpu_mpi} with minimal changes to the single-node examples.

\cprotect\subsection{\verb+lightning.kokkos+}
Given that the implementation of Algs.~\ref{alg:generic_1q_gate} and \ref{alg:controlled_1q_gate} easily avail of parallelization for the coefficient interactions, we also explored taking the same definitions and writing them using a performance-portability framework. For this, we implemented all operations using Kokkos~\cite{kokkos-1,kokkos-2}, given the direct support for OpenMP, CUDA, ROCm and SYCL, allowing us to target all backend architectures we aim to support with PennyLane.

To date, we have validated \verb+lightning.kokkos+ using the OpenMP backend across a wide variety of x86\_64, ARM, and PowerPC devices; the CUDA backend using Pascal era through to Ampere devices; and the ROCm backend for AMD Instinct devices in both the MI100 and MI200 series. As with the other devices, \verb+lightning.kokkos+ supports all operations and observables, allowing us to seamlessly change the device backend from the Python layer to target the available system hardware. When aiming to use PennyLane on any CPU or GPU device not supported by \verb+lightning.qubit+ or \verb+lightning.gpu+, or in scenarios with few observables running on systems with large numbers of CPU cores, \texttt{lightning.kokkos} can offer the best performance for workloads in these scenarios.

Given that Kokkos supports at most 1 accelerator device per process, we also have preliminary support for the mpi4py \newline\texttt{concurrent.futures} interface, allowing batching of the forward execution and adjoint differentiation with the expectation values calculations of the \texttt{qml.Hamiltonian} data-type, allowing us to distribute execution of observable terms over available resources~\cite{mpi4py_conc_fut,lightning_paper_repo2024}.

\section{Performance comparisons}\label{sec_performance}
To assess the performance of the implementations discussed in Section~\ref{sec_pl_arch}, we explore the runtime performance for a set of examples, as well as demonstrate the use of the devices on different hardware backends. All example scripts and configurations for the included data can be found at the following repository~\cite{lightning_paper_repo2024}.

\subsection{Analysis of SIMD kernel performance with \texttt{lightning.qubit}}\label{subsec:simd_perf}

To better understand the impact of intrinsic-enabled kernels on runtime performance, we microbenchmark a subset of the gate kernels. To avoid all Pythonic overheads, we showcase the gate application times by comparing the default (LM) portable gate kernels, to the AVX2 and AVX-512 kernels, all running on the same Sapphire Rapids hardware (AWS EC2 R7i.16xlarge) directly from a C++ binary. Lightning-Qubit was compiled with the \newline\texttt{-DLQ\_ENABLE\_KERNEL\_OMP=ON} option to enable gate-kernels multi-threading, and used variants of the Lightning v0.33 and v0.34 releases. All examples were compiled to use double-precision representations throughout. For an internal comparison across Lightning-Qubit gate kernels, we will consider a 30-qubit state vector using the single-qubit RX gate ($\exp(i\frac{\theta}{2}\sigma_x)$) across the range of qubit indices and gate kernels.

\begin{figure}
  \includegraphics[width=0.99\linewidth, trim={0.3cm 0.1cm 0.1cm 0},clip]{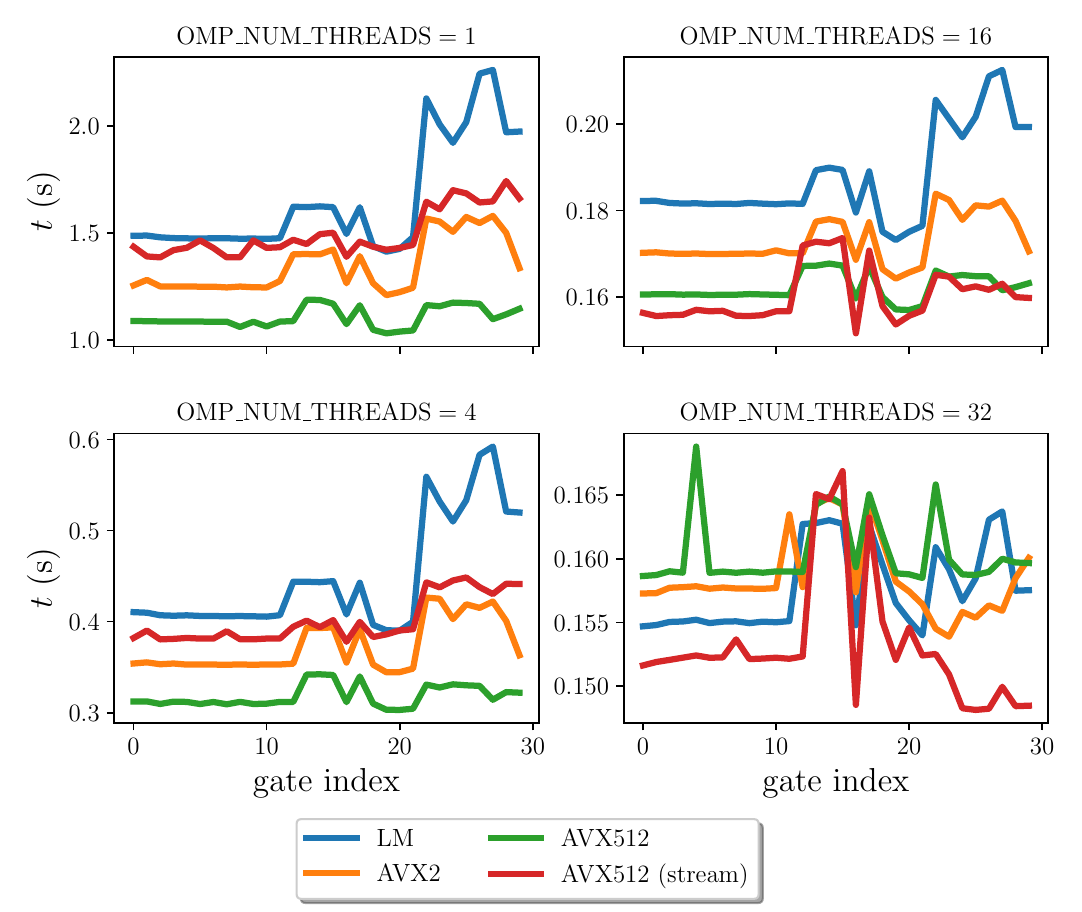}
  \caption{Performance results for the Lightning-Qubit RX gate kernels, comparing the default (LM), AVX2, AVX-512 and AVX-512+streaming kernel implementations across gate indices and OpenMP threads for a 30-qubit state vector. The kernels show improved performance from default to AVX2, and again to AVX-512 for the one and four and 16-threaded workloads, with the default LM kernels taking the lead elsewhere. The addition of streaming operations to the AVX-512 kernels shows advantage at higher thread counts, with the performance between the LM and AVX2 kernels for lower counts.}\label{fig:rx_gate_lightning_kernels}
\end{figure}

Figure~\ref{fig:rx_gate_lightning_kernels} presents an interesting landscape of performance results. The AVX-512 gate performance yields large gains over the AVX2 and default kernels (LM), up to at least 16 threads on the 32 physical core socket. Of note is the saturation of the intrinsic kernel performance at the higher thread counts, wherein the default multithreaded kernel application gains the lead. Through a \verb+perf c2c+~\cite{perf_c2c} analysis we discovered this to be in part due to cache-eviction, as an increased number of load \& store operations appeared at the higher thread counts. A mitigating strategy for higher thread counts here was employed using streaming operations (converting \verb+_mm512_store_pd+ calls to \verb+_mm512_stream_pd+) to enable bypassing the cache update upon kernel exit. This shows benefit at high-thread counts, with the performance at lower counts somewhere between the AVX2 and default kernel implementations.

\subsubsection{Comparison with other simulators}

\begin{figure}
  \includegraphics[width=0.99\linewidth, trim=0.25cm 0.5cm 0 0, clip]{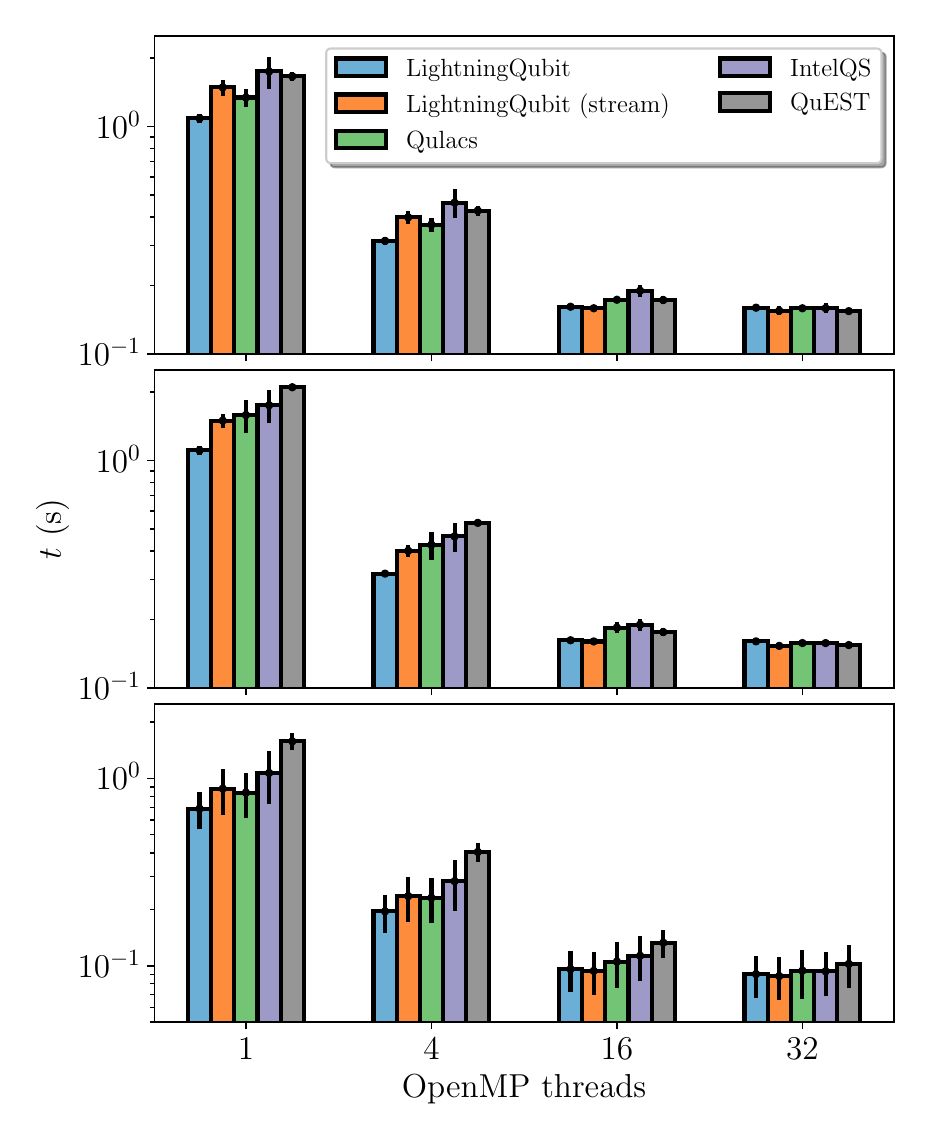}
\caption{A comparison of the runtime averaged gate performance for Hadamard (top), RX (middle), and CNOT (bottom) across a variety of high-performance quantum simulator frameworks for a 30-qubit state vector. The AVX-512 kernels for Lightning-Qubit show an advantage for a single thread, as well as many of the multithreaded regimes, with the advantage tapering off for the highest count. For higher thread-counts, enabling the streaming AVX-512 operations recovers the performance advantage.}\label{fig:runtime_all_simulators}
\vspace{-2.0em}
\end{figure}

For comparison, we have selected some commonly used state vector simulators and compared the runtime performance across a series of gates.
We have run the benchmark over every available index for single-qubit gates, and every gate index pairing for two-qubit gates. We have built each respective framework in a reproducible way and in the most performant manner recommended for each framework, including enabling targeted autovectorisation where possible (see~\cite{lightning_paper_repo2024} for further details). The performance evaluation was performed to investigate the gate runtime in the regime for intermediate scale quantum (ISQ) algorithm development, which we imagine starting to take place at 30-qubits. 
While the runtime landscapes can differ across devices, distilling a single value from the data required averaging all runtime numbers over the number of runs, with each backend and OpenMP thread count used as indicators of scaling behaviour. These results can be seen in Fig.~\ref{fig:runtime_all_simulators} for the RX, Hadamard, and CNOT gates.

For the fairest comparison, we built all binaries with well-optimized compiler flags and ensured any kernels that could benefit from AVX-512 were also given feature flags to enable support. As can be seen, the Lightning-Qubit AVX-512 kernels offer the best single-threaded performance and the lowest runtime variance over all examined indices. For the highest thread count, the AVX-512 kernels lose their advantage, as observed with Fig.~\ref{fig:rx_gate_lightning_kernels}, though the range of value differences in this regime is on the order of $10^{-3}$ seconds. As \verb+lightning.qubit+ is primarily built to target workloads that involve quantum circuit gradients, the single-threaded performance is of the utmost importance as each observable in the gradient pipeline is spawned as a separate OpenMP thread, where the evaluations happen concurrently. As no other framework supports this level of parallelism, we can achieve both algorithmic gains through use of the adjoint differentiation method over parameter-shift and finite-difference methods, as well as improvements through threading at this level. As such, comparisons between the adjoint differentiation supported gradient pipeline and the non-supported frameworks would not be a fair comparison in this regime due to the quadratic parameter execution complexity of hardware-efficient methods, versus the linear complexity of parametric gate applications when using the adjoint-gradient evaluations with simulators~\cite{jones2020efficient,lee2021adjoint}.

\subsection{Single-node workload: the variational quantum eigensolver}\label{sub:vqe_1node}

A well examined application of current generation hybrid quantum workloads is material modeling with variational quantum eigensolvers (VQE)~\cite{Peruzzo2014}.
Given the economical motivations behind material modelling (alloys, batteries, photovoltaics, semiconductors, etc.) VQE is a potential avenue to begin exploring this space, and so we provide some benchmarks using the Lightning devices. Let's start with a quick look at Lst.~\ref{lst:vqe_h2o}.

\begin{lstlisting}[label={lst:vqe_h2o}, style=CodeStyle, language=Python, caption={Total energy of an \ch{H2O} molecule},  captionpos=b]
data = qml.data.load("qchem", molname="H2O", basis="STO-3G")[0] # import H2O data
ham, wires = data.hamiltonian, data.hamiltonian.wires
hf, n_e = data.hf_state, data.molecule.n_electrons
dev = qml.device("lightning.qubit", wires=len(wires))
sngl, dbl = qml.qchem.excitations(n_e, len(wires))

@qml.qnode(dev, diff_method="adjoint")
def energy(weights):
    qml.AllSinglesDoubles(weights, wires, hf, sngl, dbl)
    return qml.expval(ham)
params = qml.numpy.random.normal(0, np.pi, len(sngl)+len(dbl))
print(f"total energy = {energy(params)}")
\end{lstlisting}

PennyLane allows to conveniently load quantum datasets for exploration~\cite{Utkarsh2023Chemistry}.
Here we use PennyLane's \texttt{qml.AllSinglesDoubles} template which prepares a correlated state by applying all single (rotations in the subspace $\{\vert 01 \rangle, \vert 10 \rangle\}$), and double (rotations in the subspace $\{\vert 0011 \rangle, \vert 1100 \rangle\}$) parametric excitation operations to the initial Hartree-Fock state.
We can compute the total energy with a single execution of the circuit, with evaluation of the expectation value relative to the given system's Hamiltonian.
In practice, we seek to minimize the total energy so we also want the gradient of the energy with respect to the \textit{ansatz} parameters.
In Lightning, this is most efficiently done with the adjoint differentiation method, set via \texttt{diff\_method="adjoint"} at the QNode level.

\begin{figure}[h!]
  \includegraphics[width=0.99\linewidth, trim=0.25cm 0.55cm 1.5cm 0.75cm, clip]{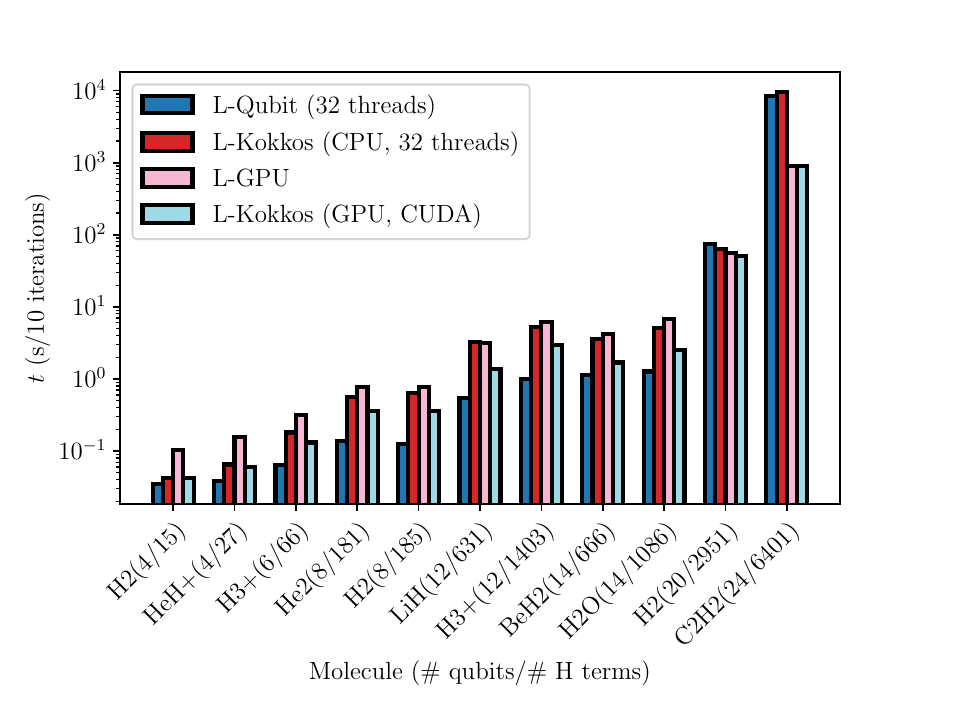}
  \caption{The time to compute the total energy and its gradient 10 times for various molecules. The number of qubits and the number of terms in the molecular Hamiltonian for each molecule are shown in parentheses.}
  \vspace{-1.em}
  \label{fig:VQE_plugins_time_vs_molecules}
\end{figure}

We compute the total energies of various molecules and their gradients 10 times with each Lightning plugin and report the results in Fig.~\ref{fig:VQE_plugins_time_vs_molecules}.
For most molecules, the electrons are captured with a minimal STO-3G basis.
Several molecules, \ch{He2} and \ch{H3+}, have a 6-31G basis, and \ch{H2} is also simulated with a CC-PVDZ basis (in addition to STO-3G and 6-31G).
Lightning-Qubit and Lightning-Kokkos (OpenMP) are run on a 32-core AMD Epyc (Milan) processor, and Lightning-GPU and Lightning-Kokkos (\texttt{-DKokkos\_ENABLE\_CUDA=on}) are run on a 40 GB NVIDIA A100 GPU, all of which are run through the ISAIC HPC platform. These runs were performed with the v0.34.0 PennyLane release. Given the Milan architecture, we expect Lightning-Qubit to auto-dispatch gate-calls to the AVX2 kernel implementations.

Lightning-Qubit and Lightning-GPU can use observable batching to parallelize parts of the gradient evaluation. In this instance, Lightning-Qubit uses up to 32 OpenMP threads. Lightning-GPU uses a single GPU for serial observable executions, allowing a fair comparison against Lightning-Kokkos which is compiled with Kokkos' CUDA backend.

For small Hamiltonians (less than 20 qubits), overheads dominate and Lightning-Qubit is the fastest plugin.
Simulating \ch{H2} with a CC-PVDZ basis requires a 20 qubit circuit and has a 2951-term Hamiltonian.
At that point, GPUs start to shine with a factor of 2 improvement in runtime performance.
Note how Lightning-GPU has larger overheads than Lightning-Kokkos for all but the largest system (\ch{C2H2}), which still shows for \ch{H2} with a CC-PVDZ basis.  
\ch{C2H2} has a 24 qubit Hamiltonian with 6401 terms, where Lightning-GPU is then on par with Lightning-Kokkos and both are around 7 times faster than Lightning-Qubit threading on a Sapphire Rapids CPU.
Beyond this scale, GPUs are required to put in the iterations necessary to minimize the total energy.

Given access to wide varieties of different hardware platforms, as well as quantum problems of interest, having a suite of backends that can be tailored to give the best outcomes for each problem is of paramount importance, and a main focus on our support across both tailored and performance-portable implementations.

\section{Quantum circuit execution across HPC systems}\label{sec_heterogeneous}

As the development of quantum computing hardware will largely be tied to, and cooperate with, advances in HPC systems for some time, we anticipate the need for exploring how quantum software environments can work effectively and interact with heterogeneous execution environments, expanding from the CPU \& GPU paradigms, to also support QPUs. With the series of supported devices and different architectures we can target, we see PennyLane \& Lightning as useful proxies to examine interactions in these environments, and additionally to interact with quantum hardware with different supports.

Through PennyLane we can set up several variations of CPU and GPU devices as proxies for multiple devices interacting in environments useful to quantum computing research, at the algorithm level, the classical-quantum interoperation level, and the large-scaled workload level. In the following section, we will build and examine a selection of workloads to demonstrate the multi-node supports for Lightning backends.

Here we will explore three separate example workloads:
\begin{enumerate}[i]
\item A distributed task-based circuit cutting workload \newline(\texttt{lightning.gpu}, Ray, Perlmutter).
\item A distributed task-based variational quantum eigensolver optimization workload (\texttt{lightning.kokkos} [HIP/ROCm], \newline {MPI/mpi4py}, LUMI).
\item An MPI-distributed state vector finite-shot (sampling) circuit execution workload (\texttt{lightning.gpu}, MPI, Perlmutter).
\item An MPI-distributed state vector quantum circuit gradient workload using the adjoint gradient method (\texttt{lightning.gpu}, MPI, Perlmutter).
\end{enumerate}


\subsection{\label{subsec_task_circ_workflow}Task-based circuit execution}

One defining feature of PennyLane is the native support for quantum circuit transformations: in essence, applying circuit mappings prior to execution. This allows us the ability to define a quantum circuit of interest, and map it to explore related quantities. In this section, we will explore a task-based workload wherein a circuit of interest is too large to run on a given device (simulator or hardware), and requires pre- and post-processing to fit onto available resources.

Let us consider the QAOA MAXCUT~\cite{farhi2014quantum,farhi2019quantum,Hadfield_2019} workload demonstrated by~\cite{Lowe2023fastquantumcircuit}. A clustered graph is built and mapped onto a quantum circuit for a QAOA demonstration of a MAXCUT problem. The parameters defined by $r=3$ (three node clusters), $n=25$ (25 nodes per cluster), $k=2$ (2 edge separators), and $l=1$ (a single QAOA layer), yield a 79 qubit requirement to represent the circuit. Using the circuit-cutting transform in PennyLane allows us to decompose this problem into smaller chunks, each of which can be independently solved, followed by a postprocessing step to reconstruct the final expectation value of the Hamiltonian. For the chosen problem, we have naively expressed the circuits (i.e. avoiding efficient groupings and optimizations), yielding a total of 45711 (excluding the offset identity circuit) fragments to run within a qubit range of 27-29.

To efficiently evaluate this large number of circuits, we use \verb+lightning.gpu+ running on the NERSC Perlmutter system, with Ray for task orchestration. We applied two separate approaches to running the circuit fragments:
\begin{enumerate}[i]
\item Creating $g$ batchable chunks of circuits, where $g$ is the number of available GPUs, and evaluating each fragment within a chunk iteratively on a single device allocation. As the generated circuits feature a variety of qubit size requirements, this requires us to allocate an initial GPU device with sufficient memory to accommodate all range of circuits in a given chunk.
\item Scheduling each circuit to run on-demand for a given GPU device instance, which dynamically allocates the required memory per circuit --- this can allow us to avoid the overheads of running smaller circuits on a larger state-vector representation, at the expense of dynamically allocating sufficient memory per circuit.
\end{enumerate}

\begin{figure}
  \centering
  \includegraphics[width=0.9\linewidth, trim={0.3cm 0.35cm 0cm 0.0cm},clip]{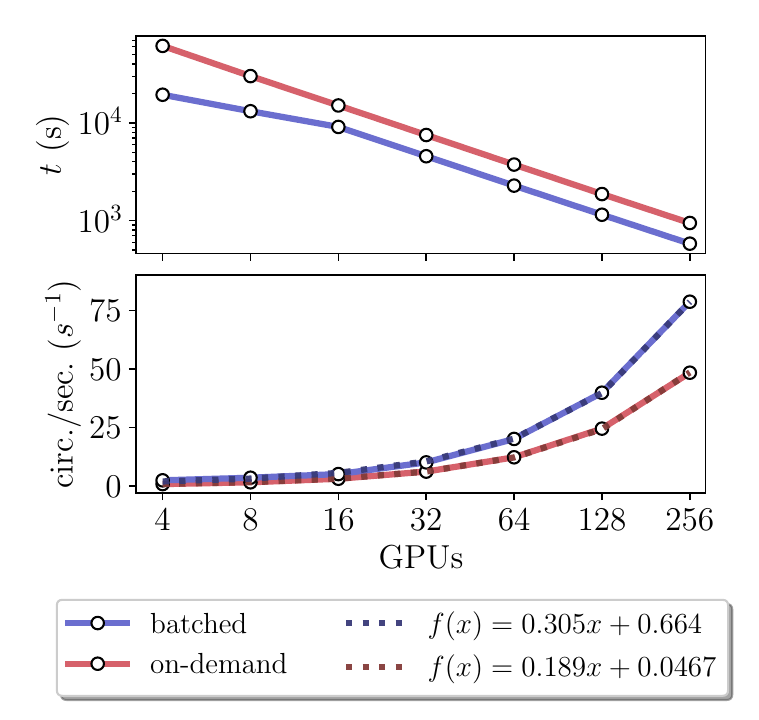}
  \caption{Circuit executions across GPUs for the given QAOA cut-circuit example. Two execution strategies were employed for the 45711 circuits, which showed very different scaling behaviour over the GPU counts (circuit execution time, top) as well as aggregated circuit throughput (bottom).} 
  \vspace{-1em}
  \label{fig:qcut_scaling}
\end{figure}

Comparing the scaling behaviour from 1 to 64 nodes (4 to 256 GPUs) with the above strategies shows us
Fig.~\ref{fig:qcut_scaling}. We can see that the on-demand scheduling seems to have significantly worse performance than the batched execution across the range of examined GPUs. For the batched execution, a gap opens up at lower nodes, but closely linearises as the GPU count increases. In this case, the workload enabled us to explore useful heuristics for the given workload type, and choose the best approach for similuating similar workloads. Reframing the data as in Fig.~\ref{fig:qcut_scaling} (bottom) also provided a rough metric to determine the GPU counts required for an average number of circuit executions per second, with the fixed number of circuits determined from the given circuit cutting example. In this case, we can see that the batched execution allows us to hit a maximum throughput of approximately 80 circuits per second at the 256 GPU scale, for the the mixed 27 to 29 qubit set of circuits, with line-fits added to help with fine-tuning the resources for further workloads.

Of course, smarter scheduling strategies can be employed, such as defining a Ray actor pool to maintain resources between dynamical runs, splitting the GPUs using Ray's \verb+split_gpu+ argument for smaller chunks to be grouped together and packed to reduce I/O overheads, allowing the unused CPU cores to take a chunk of the outstanding work to compliment the GPU resources, and even defining a reusable memory pool per GPU device, but these were not considered as part of this work.

\subsection{\label{subsec_mpi_task_circ_workflow}Batched VQE optimization}
Following on from the evaluations on single-node VQE problems in Sec.\ref{sub:vqe_1node}, going to larger sized Hamiltonians can be costly in compute and memory resources. Considering \ch{C2H4}, without any clever representations or grouping to reduce the qubit count or term count~\cite{PhysRevA.105.062452, doi:10.1021/acs.jctc.0c00008, 10.1063/1.5141458}, yields a 28-qubit \texttt{AllSinglesDoubles} \textit{ansatz} circuit with over 3000 gates, and with close to 9000 Pauli words in the expectation value Hamiltonian. As each Hamiltonian term can be independently evaluated and reduced over, we can split building the Jacobian of these large Hamiltonians over a variety of resources. In this instance, we consider using \texttt{lightning.kokkos} with the ROCm (HIP) backend, and parallelize this on LUMI using mpi4py's \texttt{concurrent.futures} pool-executor API to distribute the observables.

\begin{figure}
  \centering
  \includegraphics[width=0.99\linewidth, trim={0cm 0.5cm 0cm 0},clip]{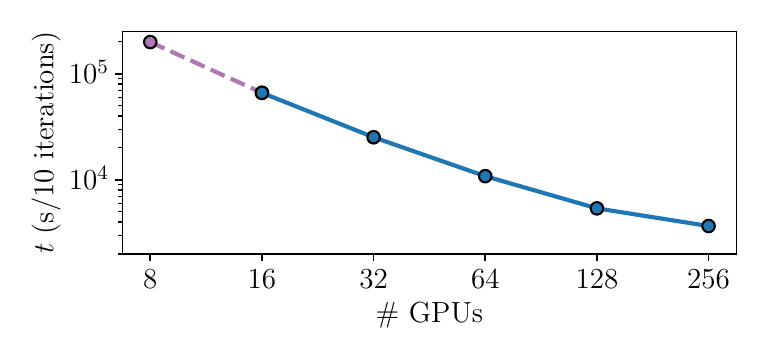}
  \caption{The gradient runtime during VQE using 10 optimization steps for \ch{C2H4} with a decomposed PennyLane AllSinglesDoubles ansatz template. A custom-build of OpenMPI 5.0, mpi4py 4.0-dev and multithreaded UCX were used to enable support for the \texttt{concurrent.futures} interface on LUMI.}
  \vspace{-1.0em}
  \label{fig:rocm_lumi}
\end{figure}

Given the large scale of the problem, we evaluate at most 10 steps of the VQE parameter optimization process, running on a variety of nodes, with the results given in Fig.~\ref{fig:rocm_lumi}. As the single node example (8 GPUs) hit the wall-clock limit at 48-hours, we estimate the runtime by sampling from distribution using the mean and standard deviation of the given 8 steps, with the rest using the exact values. The output energies for each step were confirmed as equivalent for all runs of the distribution. While algorithmic improvements can be used to reduce the required evaluations and qubit count, the larger scale and naive representation of this system provided a solid test-bed for demonstrating scaled Hamiltonian optimization workloads.

\subsection{\label{sec:dist_sv_fwd}Distributed state vector workloads}

As current real-world quantum hardware will be sample/finite-shot based, the ability to run larger-scaled examples of workloads mirroring this can be beneficial in regimes where algorithmic developments and validation is required.

Here we will explore a parametric circuit workload, composed of a PennyLane template taking inspiration from the circuit-centric classifier design~\cite{schuld_2020}, namely \verb+qml.StronglyEntanglingLayers+. For each qubit and layer in the template, we require $L\times q \times 3$ parameters, and using a range of $r=1$ we ensure all controlled interactions are with the adjacent index modulus the number of qubits, $q$. The template features gates requiring 3 parametric angles, composed of ZYZ Euler angle rotations, and CNOTs. Our goal is to distribute the execution of this circuit template over a range of nodes and qubit, while taking $10^4$ samples from the circuit.

\begin{figure}
  \centering
  \includegraphics[width=0.8\linewidth, trim={0.5cm 0.5cm 0.4cm 0.5cm},clip]{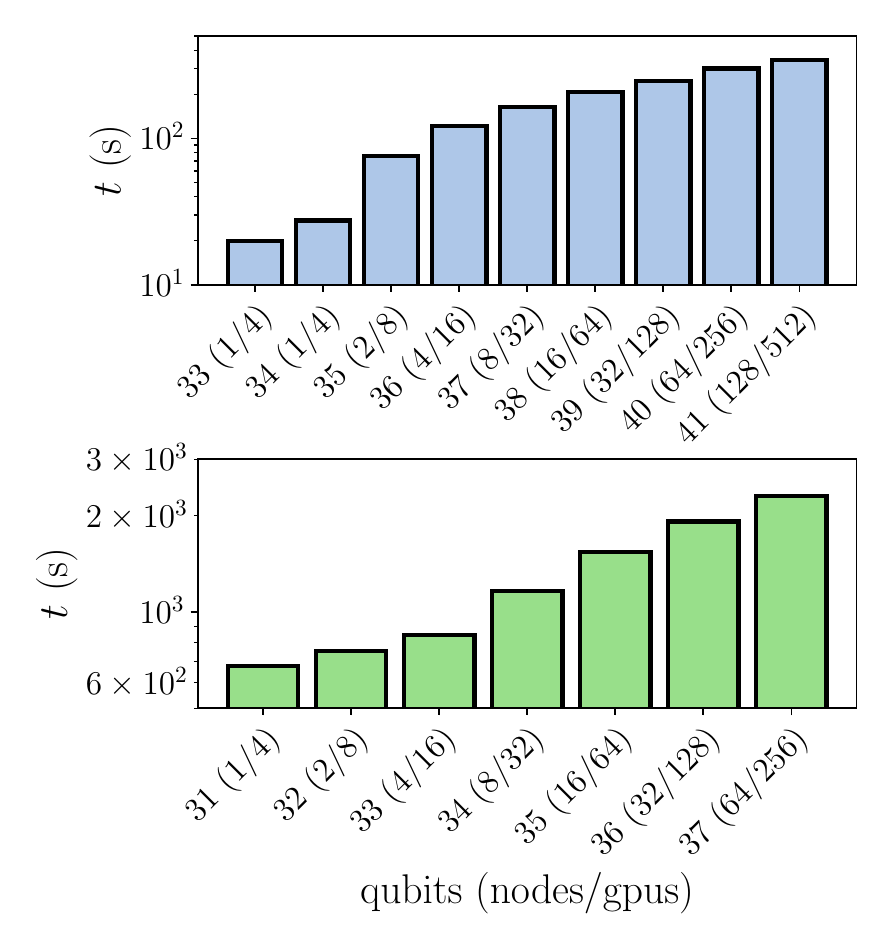}
  \caption{MPI-distributed state vector simulations of a hardware-like sampling workload (top) and an exact circuit gradient workload (bottom) running on \texttt{lightning.gpu} for the circuit-centric classifier inspired circuit using the MPI-enabled backend. These ran on Perlmutter using the 80GB A100 variant cards, atop Cray-MPICH and CUDA 11.7, as of December 2023.}
  \vspace{-1.5em}
  \label{fig:mpi_scaling}
\end{figure}

In addition, and as an extension of the above workload, we also explore the same circuit but in the context of an exact expectation value calculation with gradient evaluations. We modify the measurement process to include an expectation value output for a Pauli-Z observable for each qubit in the circuit. The runtimes of both workloads are given by Fig.~\ref{fig:mpi_scaling}, showing a weak scaling efficiency of 8\% and 30\% for the sampling and gradient workloads, and for the largest GPU counts (256/512) respectively. As previously mentioned, the adjoint method is employed to build and return the Jacobian matrix to PennyLane on all Lightning devices, including the distributed implementation of \texttt{lightning.gpu}. To our knowledge, \texttt{lightning.gpu} is the only distributed state vector implementation supporting this gradient method.

\section{Summary and conclusions}\label{sec_summary}

In this document, we have explored the suite of PennyLane's Lightning simulators. We defined our approach to gate-kernel design, providing a template to build single, controlled, and multi-qubit gate kernels. We discussed the implementation of these algorithms for \texttt{lightning.qubit}, with an investigation of kernel performance improvements using SIMD intrinsics. For gate applications at 30 qubits, we see the advantage of using AVX-512, and AVX2, showcasing an improved performance of the gate application time relative to other packages for a subset of gates, for both single-threaded and multi-threaded implementations.

To showcase the kernels in practice, we examined a variety of workloads across different scales, including the variational quantum eigensolver across \texttt{lightning.qubit}, \texttt{lightning.gpu} and \texttt{lightning.kokkos} on a single node for a variety of molecules. For smaller systems, the CPU-backed simulators showed an advantage, due in part to overheads incurred at the GPU layer. For larger systems, the GPU backends showed an increased performance advantage relative to the CPU, and become important as workloads go beyond 20 qubits.

We then explored the use of the simulators in distributed execution environments, for both task-based and distributed model-parallel workloads. In this regime, we showed support for a variety of hardware accelerator architectures, natively supported quantum circuit gradients, and provided workload runtime numbers helpful for resource estimation. Providing estimates for the resource requirements of multiple quantum workloads used in research is important to understand resource requests and allocations, so we provided some guidelines on determining approximate runtime estimations for a variety of large-scale problems.

From here, we see the Lightning suite through PennyLane as a fully-featured quantum programming toolkit with native support for HPC workloads across multithreaded CPU, batched execution, NVIDIA CUDA, AMD ROCm, and MPI-distributed environments, with the added support for native and efficient quantum circuit gradient evaluations. This wide variety of supported backends execution environemtns allows researchers to target the problem to their available hardware access.

As a follow-on from the implementations shown, our goal is to better explore the native integration of VJP and JVP methods across the various backends, a feature that recently landed in PennyLane, allowing more efficient evaluations of cost-functions through the various supported auto-differentiation frameworks. In addition, with HPC platforms being a necessity for exploring quantum research problems, offering the widest support for CPU and GPU platforms will be of paramount importance to quantum software frameworks until the widespread adoption of error-corrected quantum hardware, and even thereafter.

\subsection{Acknowledgements}
This work used resources from the National Energy Research Scientific Computing Center (NERSC), a U.S. Department of
Energy Office of Science User Facility located at Lawrence Berkeley National Laboratory, operated under Contract No. DE-AC02-05CH11231 using NERSC award DDR-ERCAP0022246 under the QIS@Perlmutter program.

The authors also wish to acknowledge the support provided by the LUMI facility for providing access to the platform for evaluations of \texttt{lightning.kokkos}. In addition, we acknowledge access and support provided by ISAIC for the use of their CPU and GPU resources during performance evaluations. Finally, we also acknowledge AWS for their continued support of PennyLane through both Amazon EC2 and Amazon BraKet.

\medskip
\bibliography{refs}

\clearpage
\onecolumn
\appendix

\section{Installation instructions}
\label{app:install}


\subsection{PyPI}

PyPI is the official repository for \href{https://pypi.org/project/PennyLane/}{PennyLane} and \href{https://pypi.org/project/PennyLane-Lightning/}{Lightning}'s compiled binaries, and installation through \texttt{pip} is recommended when it is available.
Lightning can be installed simply as in Listing~\ref{lst:pipvenv}.

\begin{minipage}{\linewidth}
\begin{lstlisting}[label={lst:pipvenv}, style=CodeStyle, language=bash, caption={\texttt{pip} install Lightning},captionpos=b]
python -m venv venv
source venv/bin/activate

# Install/upgrade PennyLane and `lightning.qubit`
# PyPI pennylane-lightning package implicitly installed.
pip install pennylane --upgrade

# Explicitly install PennyLane Lightning Core utilities and `lightning.qubit`
# PyPI pennylane package will also be implicitly installed.
pip install pennylane-lightning
\end{lstlisting}
\end{minipage}

Similarly, one may install \href{https://pypi.org/project/PennyLane-Lightning-GPU/}{Lightning-GPU} and \href{https://pypi.org/project/PennyLane-Lightning-Kokkos/}{Lightning-Kokkos} (OpenMP) as follows:

\begin{minipage}{\linewidth}
\begin{lstlisting}[label={lst:pipvenv_gpu}, style=CodeStyle, language=bash, caption={\texttt{pip} install Lightning-GPU/Kokkos},captionpos=b]
python -m venv venv
source venv/bin/activate

# Install/upgrade PennyLane
pip install pennylane --upgrade

# Single-node `lightning.gpu` installation
pip install pennylane-lightning[gpu]

# Choose Opt 1 or 2, depending on your CUDA version.
## Opt 1: Install CUDA 12 for Lightning >=v0.35 
pip install custatevec_cu12 
## Opt 2: Install CUDA 11 for Lightning <=v0.34 
pip install custatevec_cu11 nvidia-cuda-runtime-cu11 nvidia-cusparse-cu11 nvidia-cublas-cu11

# OpenMP lightning.kokkos variant
pip install pennylane-lightning[kokkos]
\end{lstlisting}
\end{minipage}

\subsection{Conda}

Conda-Forge is an organization maintaining a large database of conda recipes.
A so-called feedstock exists for each Lightning plugin, which enables installing \href{https://anaconda.org/conda-forge/pennylane}{PennyLane} and Lightning in Conda environments.
Lightning can be installed as in Listing~\ref{lst:condaenv}, where you may pick the Python version of your liking.

\begin{minipage}{\linewidth}
\begin{lstlisting}[label={lst:condaenv}, style=CodeStyle, language=bash, caption={\texttt{conda} install Lightning},captionpos=b]
conda create -n lightning python=3.11 pennylane
conda activate lightning
\end{lstlisting}
\end{minipage}

Mind, Lightning and PennyLane have a circular dependency, so do not try to install \texttt{pennylane-lightning} with Conda.
Simply install \texttt{pennylane} instead and Lightning will come with it.
To install \href{https://anaconda.org/conda-forge/pennylane-lightning-gpu}{Lightning-GPU} or \href{https://anaconda.org/conda-forge/pennylane-lightning-kokkos}{Lightning-Kokkos} (OpenMP) however, just replace \texttt{pennylane} in the above command by \texttt{pennylane-lightning-gpu} or \texttt{pennylane-lightning-kokkos}.

\subsection{DockerHub}

If for one reason or another PyPI and Conda are not reliable options, it is possible to use Lightning in a Docker container.
The PennyLaneAI \href{https://hub.docker.com/r/pennylaneai/pennylane}{DockerHub} account hosts five different images of the Lightning plugins, starting at \texttt{v0.32.0}.

\begin{itemize}
\item \texttt{lightning-qubit}
\item \texttt{lightning-gpu}
\item \texttt{lightning-kokkos-openmp}: this version of Lightning-Kokkos is built with Kokkos' OpenMP backend and is essentially the one found on PyPI and Conda.
\item \texttt{lightning-kokkos-cuda}: this version of Lightning-Kokkos is built with Kokkos' CUDA backend and can be executed on NVIDIA GPUs.
\item \texttt{lightning-kokkos-rocm}: this version of Lightning-Kokkos is built with Kokkos' ROCm backend and can be executed on AMD GPUs.
\end{itemize}

The procedure to create a Docker container is simple.
Taking \texttt{lightning-kokkos-openmp} as an example, one may spawn a container as in Listing~\ref{lst:docker_lk}.

\begin{minipage}{\linewidth}
\begin{lstlisting}[label={lst:docker_lk}, style=CodeStyle, language=bash, caption={Pull a Docker image and spawn a Lightning-Kokkos container.},captionpos=b]
docker pull pennylaneai/pennylane:latest-lightning-kokkos-openmp
docker run -v pwd:/io -it pennylaneai/pennylane:latest-lightning-kokkos-openmp bash
\end{lstlisting}
\end{minipage}

This will get the reader in a bash shell inside the container, with the present working directory bound as \texttt{io}.
Would one need to modify an image, a multi-stage \href{https://github.com/PennyLaneAI/pennylane-lightning/blob/v0.34.0_release/docker/Dockerfile}{Docker build} may be found in the Lightning repository in the \texttt{docker} directory.

Other container platforms compatible with Docker may also be used.
For example, Singularity usage is shown in Listing~\ref{lst:singularity_lkr}.

\begin{minipage}{\linewidth}
\begin{lstlisting}[label={lst:singularity_lkr}, style=CodeStyle, language=bash, caption={Pull a Docker image and spawn a Lightning-Kokkos container for AMD GPUs.}, captionpos=b]
singularity pull docker://pennylaneai/pennylane:latest-lightning-kokkos-rocm
singularity exec -B $HOME pennylane_latest-lightning-kokkos-rocm.sif bash
\end{lstlisting}
\end{minipage}
and Shifter usage is shown in Listing~\ref{lst:shifter_lkc}.

\begin{minipage}{\linewidth}
\begin{lstlisting}[label={lst:shifter_lkc}, style=CodeStyle, language=bash, caption={Pull a Docker image and spawn a Lightning-Kokkos container for CUDA GPUs.},captionpos=b]
shifterimg pull pennylaneai/pennylane:latest-lightning-gpu
shifter --image=pennylaneai/pennylane:latest-lightning-gpu /bin/bash
\end{lstlisting}
\end{minipage}

\subsection{Spack}

One may wish to build Lightning from source.
This can be interesting for Lightning-Kokkos for example, where performance may be tuned through the Kokkos configuration.
While this can be done by first installing Kokkos and then Lightning with their native CMake build systems, Spack is more convenient, especially if several installations coexist.
For instance, to build and install PennyLane-Lightning-Kokkos with NVIDIA-GPU support, we create and activate a Spack environment as in Listing~\ref{lst:spack_lkcuda1}.

\begin{minipage}{\linewidth}
\begin{lstlisting}[label={lst:spack_lkcuda1}, style=CodeStyle, language=bash, caption={Install Lightning-Kokkos for CUDA GPUs with Spack.},captionpos=b]
spack env create lightning-kokkos-cuda-1
spack env activate lightning-kokkos-cuda-1
spack add py-pennylane-lightning-kokkos +cuda cuda_arch=80
spack install
\end{lstlisting}
\end{minipage}

Now, supposed we would like to compare performance by tuning the Kokkos configuration.
To build and install PennyLane-Lightning-Kokkos with NVIDIA-GPU support, but with different configuration options, we create and activate a Spack environment as in Listing~\ref{lst:spack_lkcuda2}.

\begin{minipage}{\linewidth}
\begin{lstlisting}[label={lst:spack_lkcuda2}, style=CodeStyle, language=bash, caption={Install Lightning-Kokkos for CUDA GPUs with Spack.},captionpos=b]
spack env create lightning-kokkos-cuda-2
spack env activate lightning-kokkos-cuda-2
spack add kokkos +cuda cuda_arch=80 +cuda_constexpr~cuda_lambda+cuda_relocatable_device_code~cuda_uvm
spack add py-pennylane-lightning-kokkos
spack install
\end{lstlisting}
\end{minipage}

Note that in Spack \texttt{+} means the option is on and \texttt{\textasciitilde} means the option is off.
We only throw in a few options to illustrate how Kokkos is highly configurable, and how easy Spack makes it to manage configurations and installations.

\clearpage
\section{Lightning suite HPC environment set-up}
This appendix contains targeted guidelines on installing PennyLane Lightning-GPU on the NERSC Perlmutter system and Lightning-Kokkos on LUMI, for the respective workloads discussed in Section~\ref{sec_heterogeneous}.

\subsection{Lightning-GPU with MPI on Perlmutter}
Due to the recent changes in the default CUDA toolkit on Perlmutter, Lightning-GPU was updated to use CUDA 12 by default as of v0.35.0. At the time of writing, we can recommend building Lightning-GPU against the latest CUDA (12.2) Cray-MPICH (8.1.28). The following steps are subject to change over time as the system environment evolves, but as of the Lightning v0.35.0 release, the following steps will ensure a performant implementation of the distributed functionality provided by Lightning-GPU.
The instructions here were compiled from NERSC's official documentation~\cite{nerscPython2023}, and HPE-Cray's official documentation~\cite{hpeLinkBehaviour}, and should be followed in order.

\begin{minipage}{\linewidth}
\begin{lstlisting}[label={lst:lgpu_mpi_install}, style=CodeStyle, language=bash, caption={Build instructions for Lightning-GPU against Cray-MPICH.},captionpos=b]
#Load the Python module, and set up the python virtual environment
module load PrgEnv-gnu cray-mpich cudatoolkit craype-accel-nvidia80 
python -m venv lgpu_env && source lgpu_env/bin/activate

#Install mpi4py against Cray-MPICH
export MPICH_GPU_SUPPORT_ENABLED=1 # Runtime requirement from Cray-MPICH
MPICC="cc -shared" pip install --force --no-cache-dir --no-binary=mpi4py mpi4py

git clone https://github.com/PennyLaneAI/pennylane-lightning
cd pennylane-lightning && git checkout latest_release
python -m pip install custatevec_cu12 cmake ninja pennylane

#Install Lightning-Qubit, followed by Lightning-GPU. Lightning-GPU 
#will be built against Cray-MPICH and enable MPI-mediated GPU-GPU transfers
CXX=$(which CC) python -m pip install . 
CXX=$(which CC) PL_BACKEND="lightning_gpu" CMAKE_ARGS="-DENABLE_MPI=ON -DCMAKE_CXX_COMPILER=$(which CC)" python -m pip install -e . --verbose

#Ensure the MPI libraries are visible from NVIDIA custatevec
export LD_LIBRARY_PATH=$CRAY_LD_LIBRARY_PATH:LD_LIBRARY_PATH

#Allocate your nodes/GPUs and run your workload
salloc -N 2 -c 32 --qos interactive --time 0:30:00 --constraint gpu --ntasks-per-node=4 --gpus-per-task=1 --gpu-bind=none --account=<Account ID>
srun -n 8 python ./my_workload.py # 2 nodes, 4 GPUs per node
\end{lstlisting}
\end{minipage}

\subsection{Lightning-Kokkos ROCm/HIP backend on LUMI}

To successfully build a working environment with Lightning (both Lightning-Qubit and Lightning-Kokkos), we require a modern version of GCC (11.2.0 in this instance) and a recent version of the ROCm/HIP toolchain. LUMI currently has support for ROCm 5.4, which is used to build the Lightning-Kokkos GPU backend targeting the AMD MI250X GPUs. We start by ensuring the required modules are loaded, as:
\begin{lstlisting}[style=CodeStyle, language=bash]
module load rocm
module load cray-python
module load gcc/11.2.0
module load PrgEnv-cray-amd/8.4.0
\end{lstlisting}
Use of the MPI-based pool execution through the \texttt{mpi4py} \texttt{concurrent.futures} interface~\cite{mpi4py_conc_fut} requires support from a recent MPI library. As Cray-MPICH does not have support for dynamic process spawning~\cite{nerscPython2023Mpi4py}, we must build our own MPI variant with the required functionality. As we have a limited need to send the GPU-buffer data through the network interface via RDMA (we can serialize and send a single packet if needed using the existing bindings), we can use \texttt{mpi4py} as a means to send the required serialized circuit payload to another process that can spawn an AMD-GPU backed Lightning-Kokkos instance. For this, we cloned OpenMPI, and built this against UCX v1.15 with threads enabled (`--enable-mt`). The OpenMPI build was chosen as version 5.0, and configured with `--with-slurm` to allow us to run atop the allocated nodes from the LUMI scheduler. The \texttt{PATH} and \texttt{LD\_LIBRARY\_PATH} environment variables were updated to reflect the newly built OpenMPI and UCX libraries and ensure that the system-provided MPI was not used. We next set up a Python virtual environment and built \texttt{mpi4py} using the pre-released 4.0 version, using GCC 11.2 as the system compiler to avoid any potential linkage with the CRAY-PE libraries. Lightning-Qubit and Lightning-Kokkos+ROCm were then built and installed as:

\begin{lstlisting}[style=CodeStyle, language=bash]
git clone https://github.com/PennyLaneAI/pennylane-lightning/
cd pennylane-lightning
git checkout update/pickle_bindings
python -m pip install cmake ninja
PL_BACKEND="lightning_qubit" python -m pip install . --verbose
CXX="hipcc --gcc-toolchain=/opt/cray/pe/gcc/11.2.0/snos" CMAKE_ARGS="-DKokkos_ENABLE_HIP=ON -DKokkos_ARCH_VEGA90A=ON -DCMAKE_CXX_FLAGS='--gcc-toolchain=/opt/cray/pe/gcc/11.2.0/snos/'" PL_BACKEND="lightning_kokkos" python -m pip install . --verbose
\end{lstlisting}

With the packages successfully installed, we can run a workload through an allocation as:
\begin{lstlisting}[style=CodeStyle, language=bash]
salloc --nodes=8 --account=project_X --partition=dev-g --ntasks-per-node=8 --gpus-per-node=8 --time=00:15:00 mpirun --np 64 ./select_gpu python -m mpi4py.futures workload.py
\end{lstlisting}

Of note is the use of \texttt{salloc} in-line with the OpenMPI-provided \texttt{mpirun} to ensure the correct backend is used. The given command allocates 8 nodes, each with 8 GPUs (4 MI250X per node), and uses the \texttt{concurrent.futures} interface to handle the remote execution, where batches of observables are controlled by the user specifying the \texttt{batch\_obs=X} argument to the device, and controlling the concurrency in the forward pass and gradient pass with \texttt{export PL\_FWD\_BATCH=<num. GPUs>} and \texttt{export PL\_BWD\_BATCH=<num. GPUs>}. To ensure each remote process gets a GPU, following a similar guideline as defined by the LUMI guidelines~\cite{lumi_gpu_dist}, where we modify the provided \texttt{select\_gpu} script to work with OpenMPI-specific options as follows:

\begin{lstlisting}[style=CodeStyle, language=bash]
#!/bin/bash
export ROCR_VISIBLE_DEVICES=$OMPI_COMM_WORLD_LOCAL_RANK
export MPI4PY_RC_RECV_MPROBE=false # See https://github.com/mpi4py/mpi4py/issues/223
export OMPI_MCA_pml=ucx # Only use UCX
echo $ROCR_VISIBLE_DEVICES
exec $*
\end{lstlisting}

We next create our SLURM submission script and build our C2H4 example workload from the PennyLane datasets module, explicitly create our singles and doubles circuit, starting from a Hartree-Fock state, and distribute the execution across the GPUs. This workload currently requires the use of the PennyLane Lightning \texttt{update/pickle\_bindings} branch, which allows us to package and send the required data packets. The provided example can be modified to find the lowest energy by adjusting the optimizer, increasing the steps, and/or choosing a different starting state. However, given the scale of the problem (thousands of gates, thousands of Hamiltonian terms) as written this gives us a solid starting point to examine larger workloads. The associated scripts for all of the above can be found in the repository~\cite{lightning_paper_repo2024}.

\clearpage
\section{Runtime data for compiled binary comparisons}

The runtime data for the CNOT gate index comparison used within Fig.~\ref{fig:runtime_all_simulators} is presented below. To allow for the best range of comparisons within each figure, the maximum and minimum runtime values were used to set the heatmap limits. As such, the color scales are not comparable across binaries. To aid with this, the maximum, minimum, and average runtimes are labeled on the title of each figure for comparison. All examples here can be found in the manuscript repository in their respective directories~\cite{lightning_paper_repo2024}, along with explicit build options and configurations for each respective binary.

Of particular interest (and known from other such micro-benchmarks) is that not every pair of indices has equivalent runtime performance, likely due in part to the cache hierarchies of the CPU, and more generally data locality in memory. It can be reasoned that different CPUs may offer different run-time behaviour, depending upon the required workload. From here, we can see the difference between the Lightning-Qubit kernels (LM, AVX2, AVX-512, and AVX-512 with streaming), and overall gain some understanding of the regimes where each has advantages, as well as the comparative performance of the other packages. For the system used in these comparisons, a Sapphire-Rapids-based AWS EC2 R7i.16xlarge instance, the use of Lightning-Qubit with AVX-512 intrinsics offered the best performance benefit for the single-threaded workloads, as well as the heavily multi-threaded workloads when using the streaming intrinsics.

\subsection{Lightning-Qubit data}

All Lightning-Qubit binaries were built using branch \texttt{kernel\_omp} except the streaming examples, which used branch \texttt{bm\_stream}, with GCC (g++) 11 as the compiler in both cases. The following datasets were compiled with the \texttt{-DPLQ\_ENABLE\_OMP=1} flag to provide gate-level multithreading as opposed to gradient-level multithreading, which is enabled by default. As of writing, branch \texttt{kernel\_omp} is currently supported under the v0.35.0 release of PennyLane Lightning, with compile-time support for the streaming operations planned to follow in the v0.36.0 release.

\begin{table}[!htb]
    \begin{tabular}{c|c|c|c}
        \includegraphics[width=.24\textwidth, trim={2.5cm 0 0cm 0}, clip]{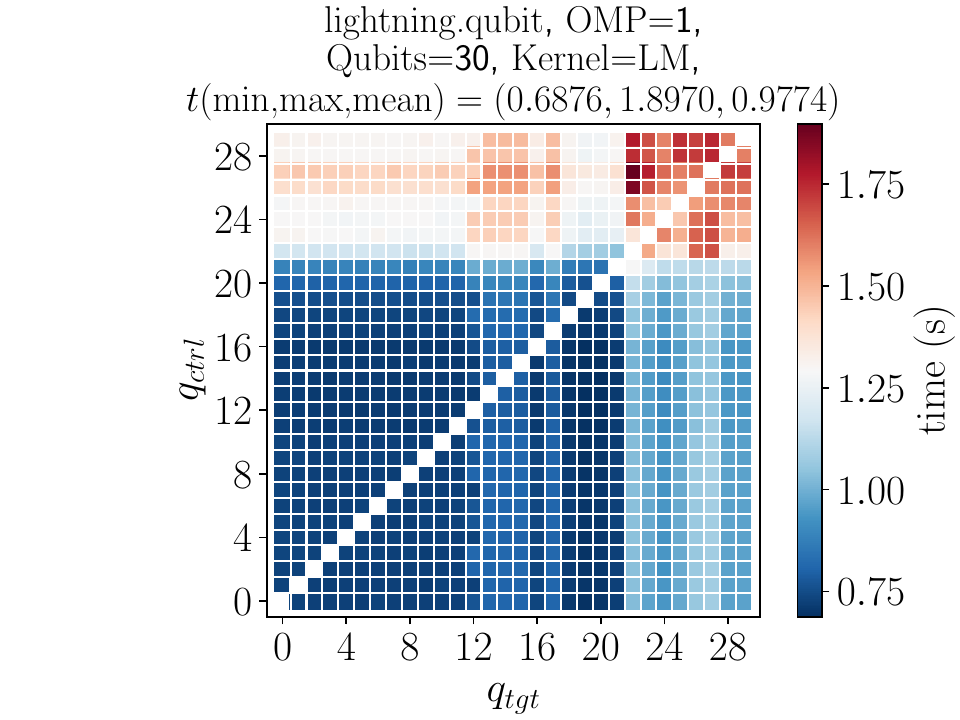} & \includegraphics[width=.24\textwidth, trim={2.5cm 0 0cm 0}, clip]{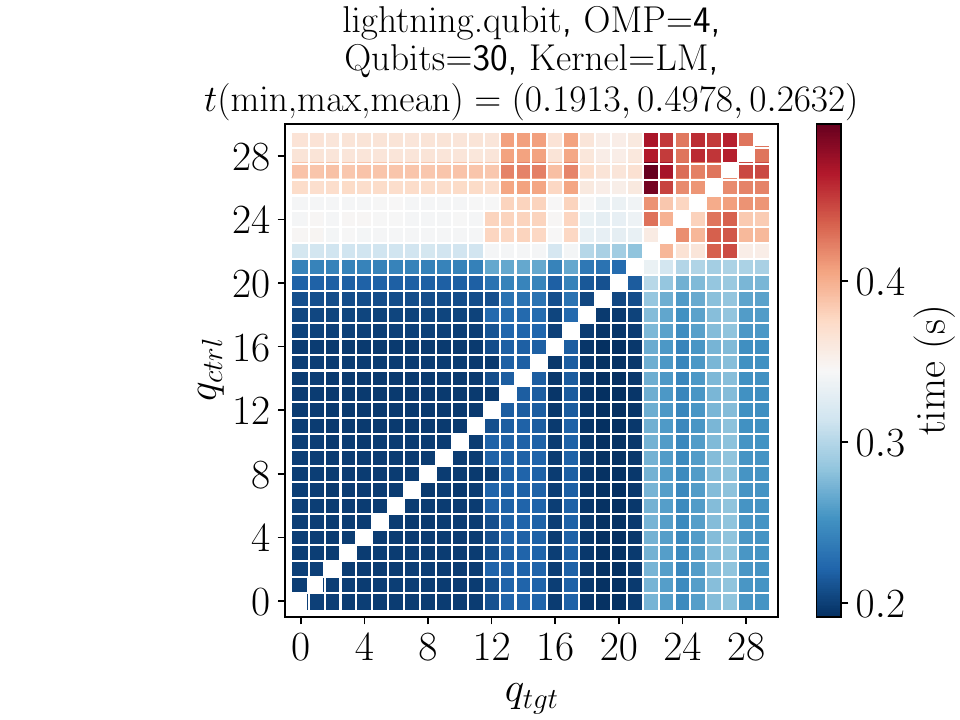} & \includegraphics[width=.24\textwidth, trim={2.5cm 0 0cm 0}, clip]{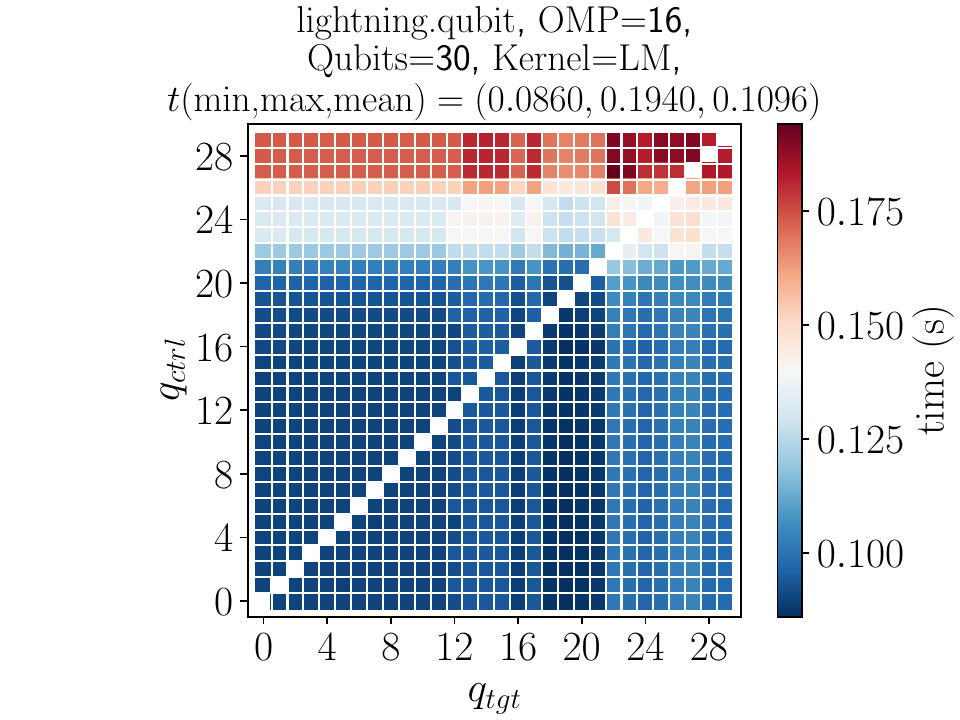} & \includegraphics[width=.24\textwidth, trim={2.5cm 0 0cm 0}, clip]{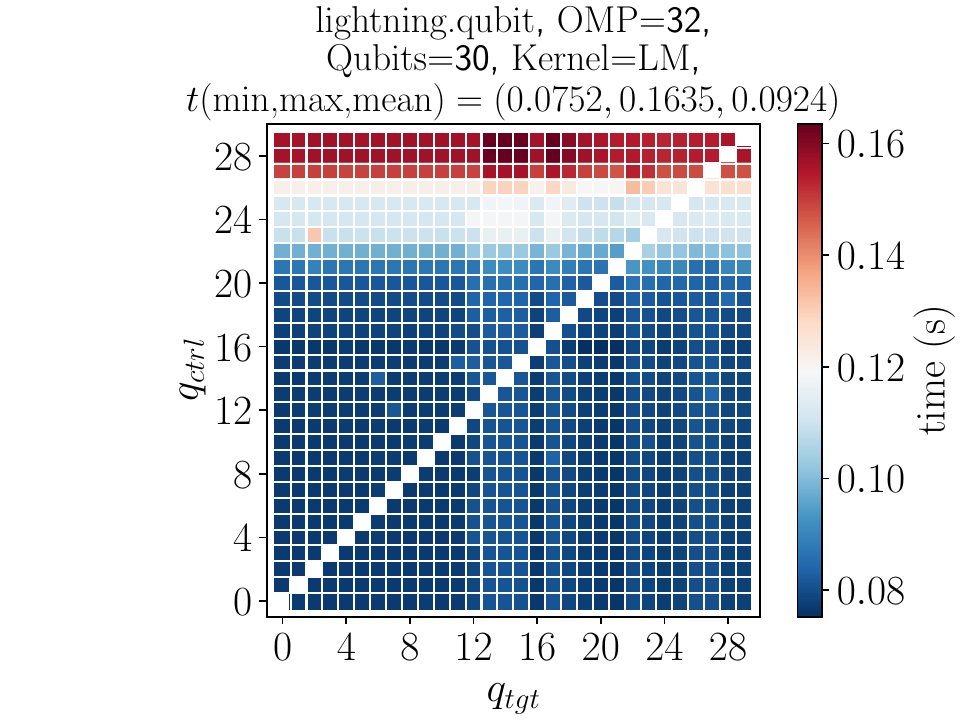}
    \end{tabular}
    \caption{Default (LM) kernel data.}
\end{table}

\begin{table}[!htb]
    \begin{tabular}{c|c|c|c}
        \includegraphics[width=.24\textwidth, trim={2.5cm 0 0cm 0}, clip]{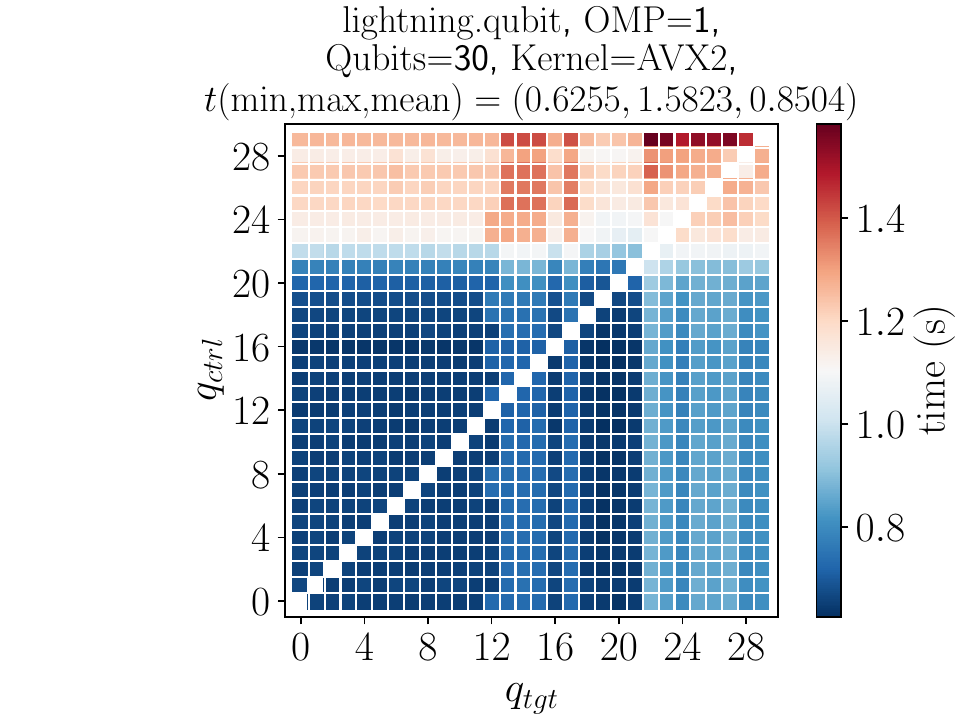} & \includegraphics[width=.24\textwidth, trim={2.5cm 0 0cm 0}, clip]{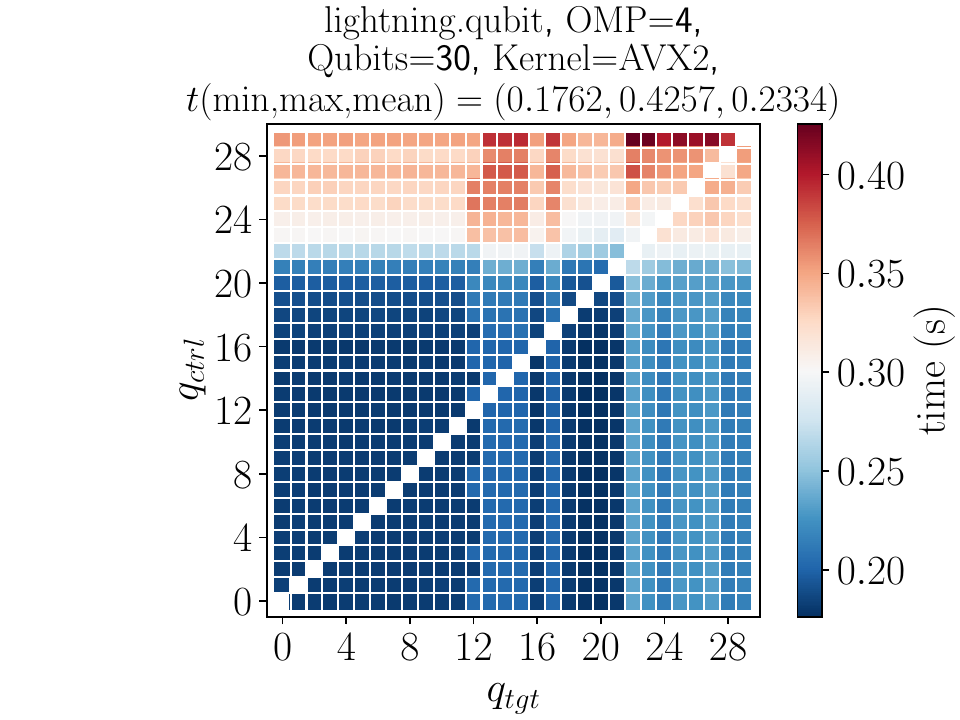} & \includegraphics[width=.24\textwidth, trim={2.5cm 0 0cm 0}, clip]{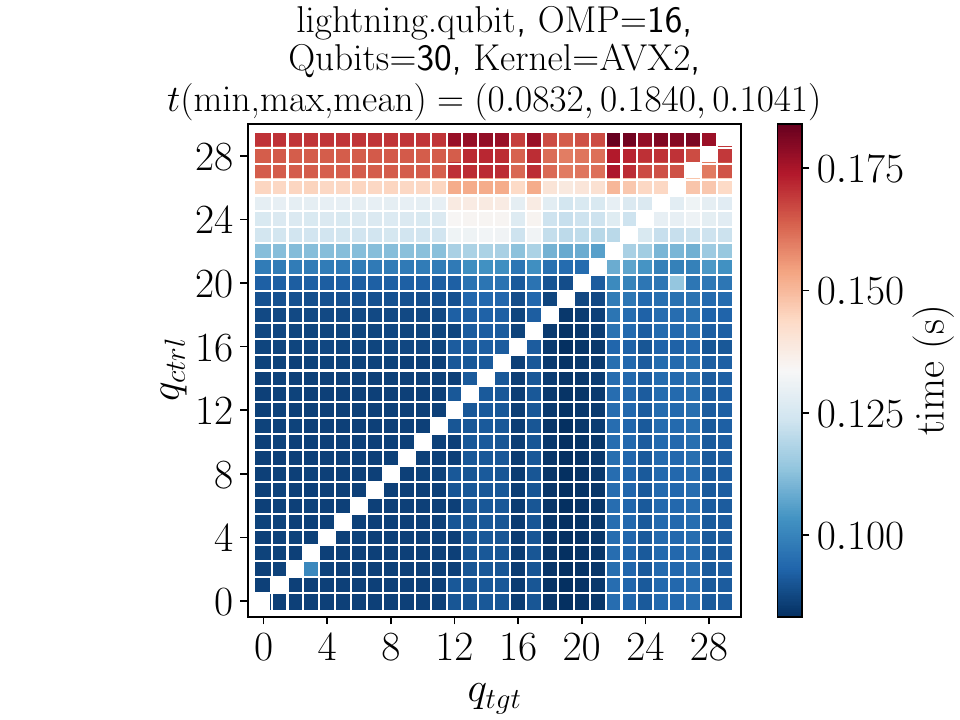} & \includegraphics[width=.24\textwidth, trim={2.5cm 0 0cm 0}, clip]{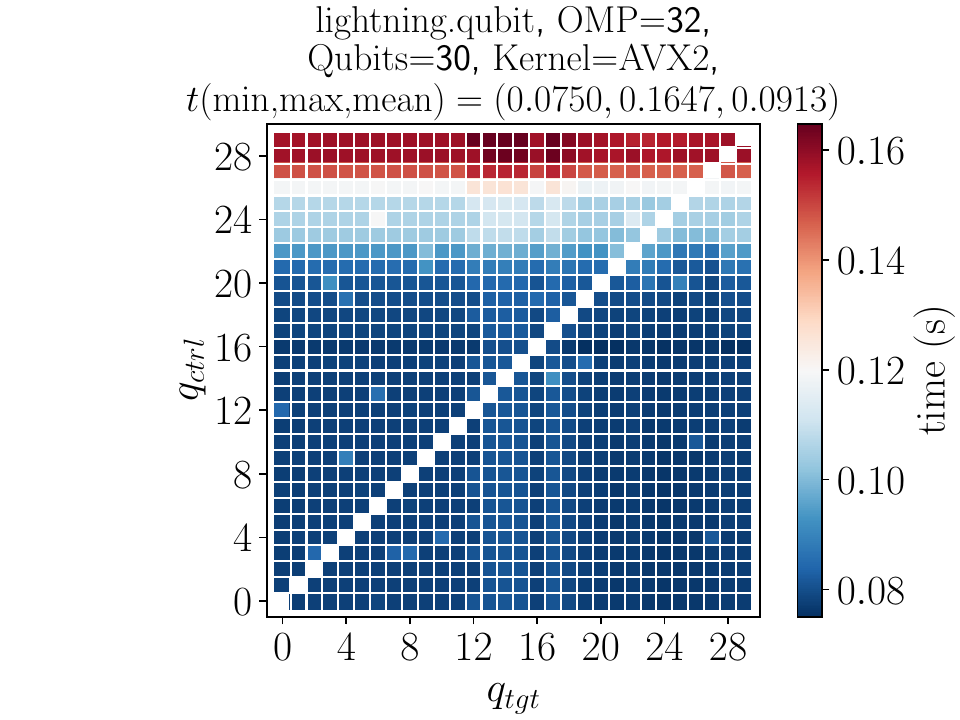}
    \end{tabular}
    \caption{AVX2 kernel data.}
\end{table}

\begin{table}[!htb]
    \begin{tabular}{c|c|c|c}
        \includegraphics[width=.24\textwidth, trim={2.5cm 0 0cm 0}, clip]{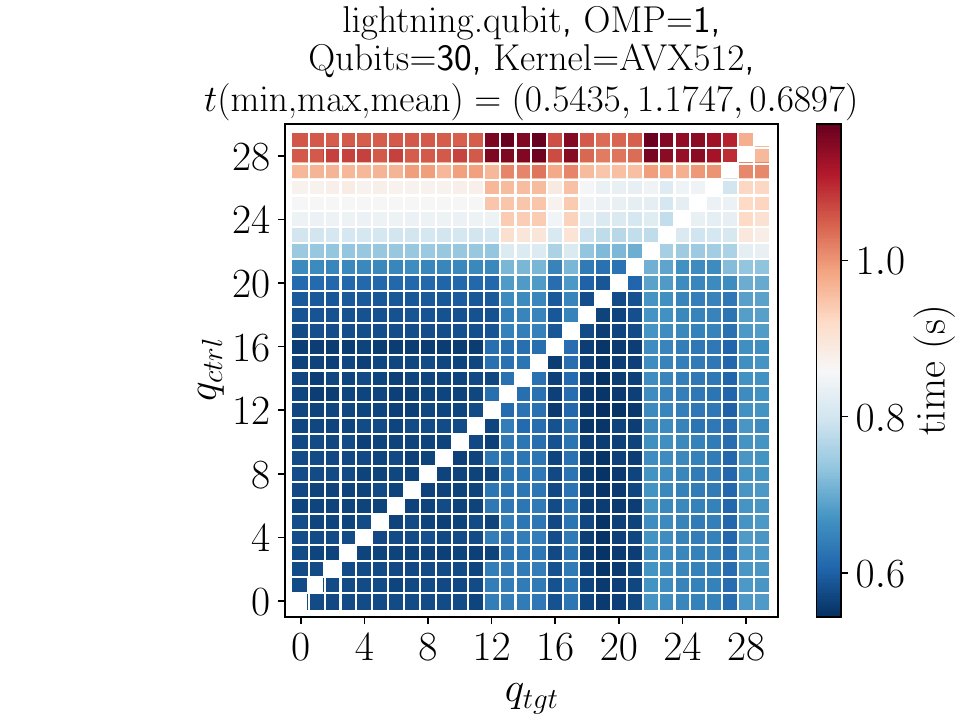} & \includegraphics[width=.24\textwidth, trim={2.5cm 0 0cm 0}, clip]{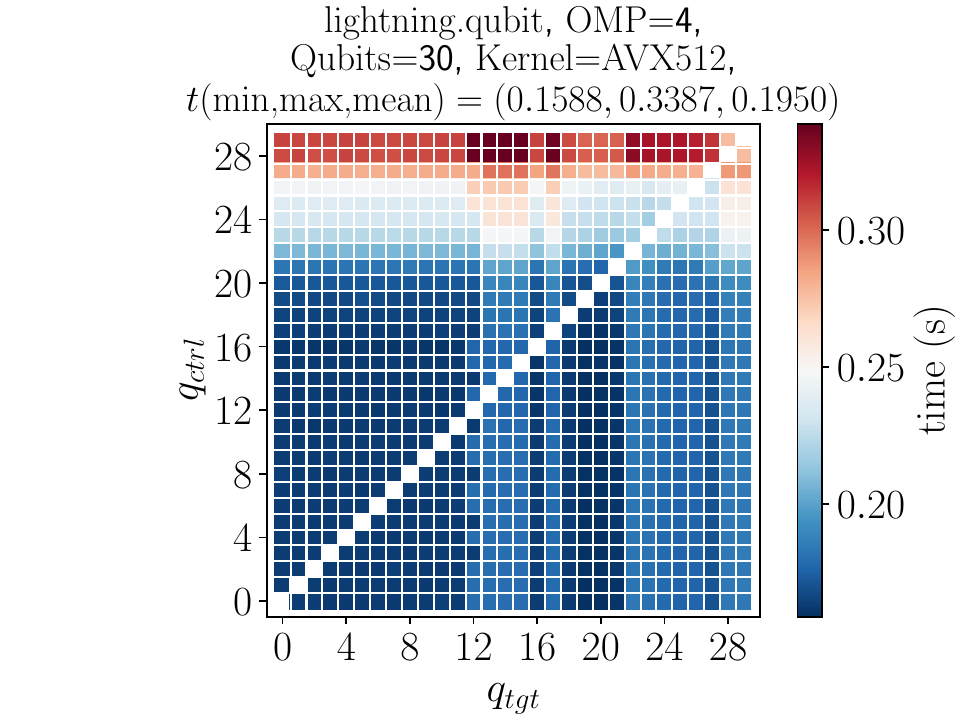} & \includegraphics[width=.24\textwidth, trim={2.5cm 0 0cm 0}, clip]{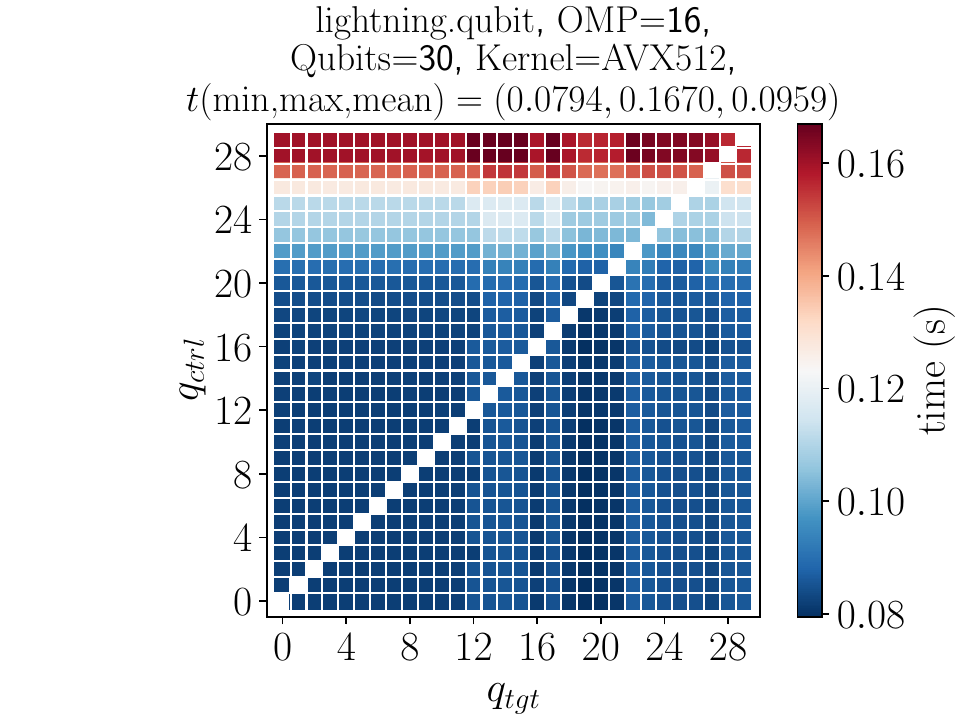} & \includegraphics[width=.24\textwidth, trim={2.5cm 0 0cm 0}, clip]{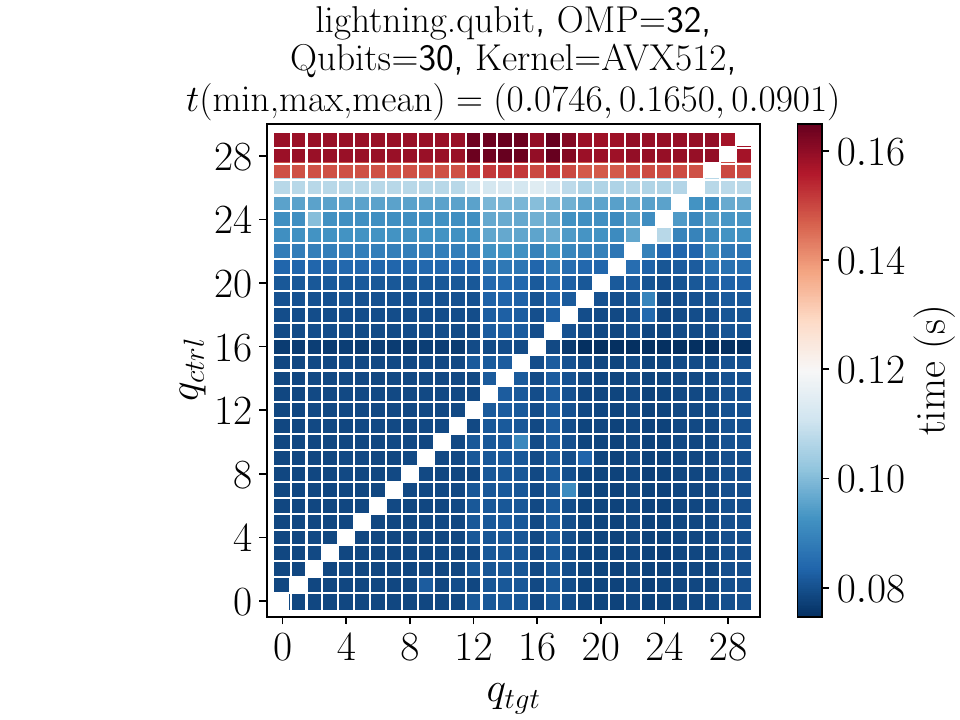}
    \end{tabular}
    \caption{AVX-512 kernel data.}
\end{table}

\begin{table}[!htb]
    \begin{tabular}{c|c|c|c}
        \includegraphics[width=.24\textwidth, trim={2.5cm 0 0cm 0}, clip]{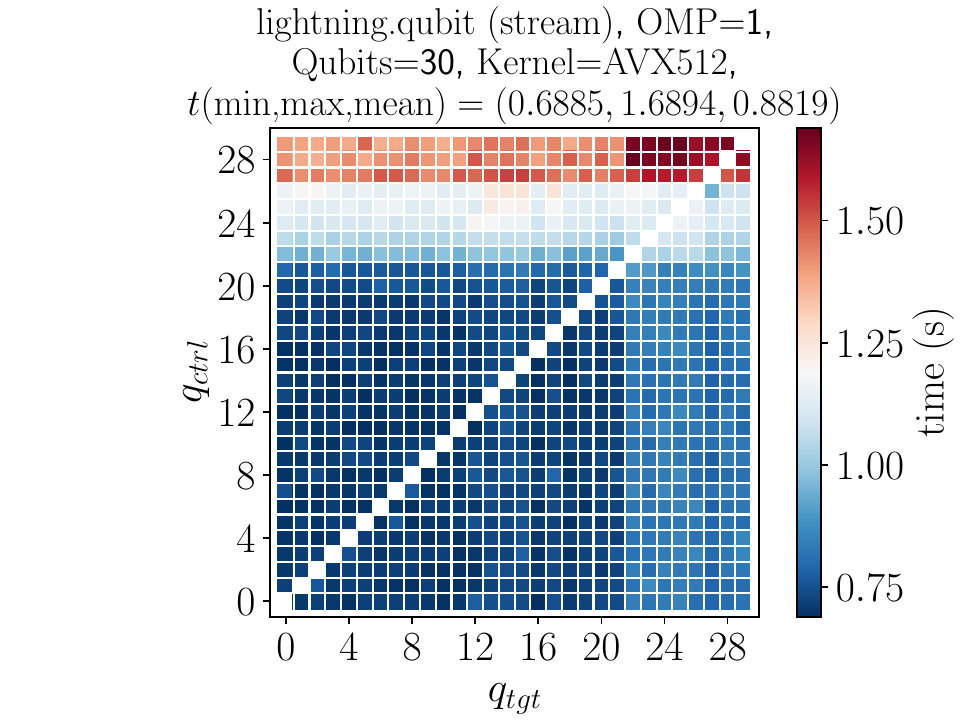} & \includegraphics[width=.24\textwidth, trim={2.5cm 0 0cm 0}, clip]{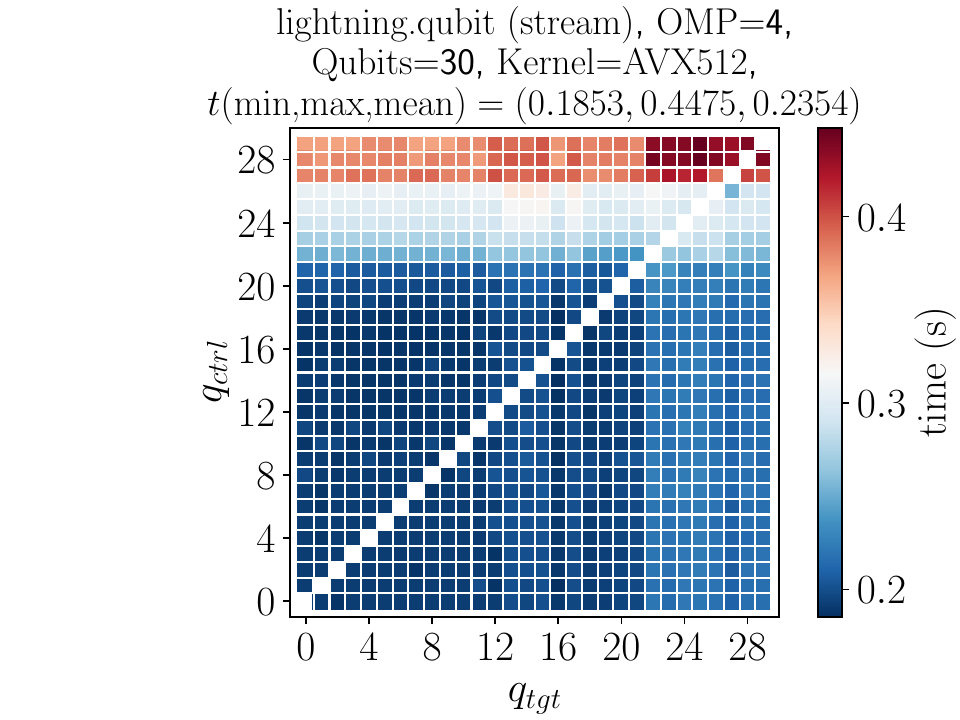} & \includegraphics[width=.24\textwidth, trim={2.5cm 0 0cm 0}, clip]{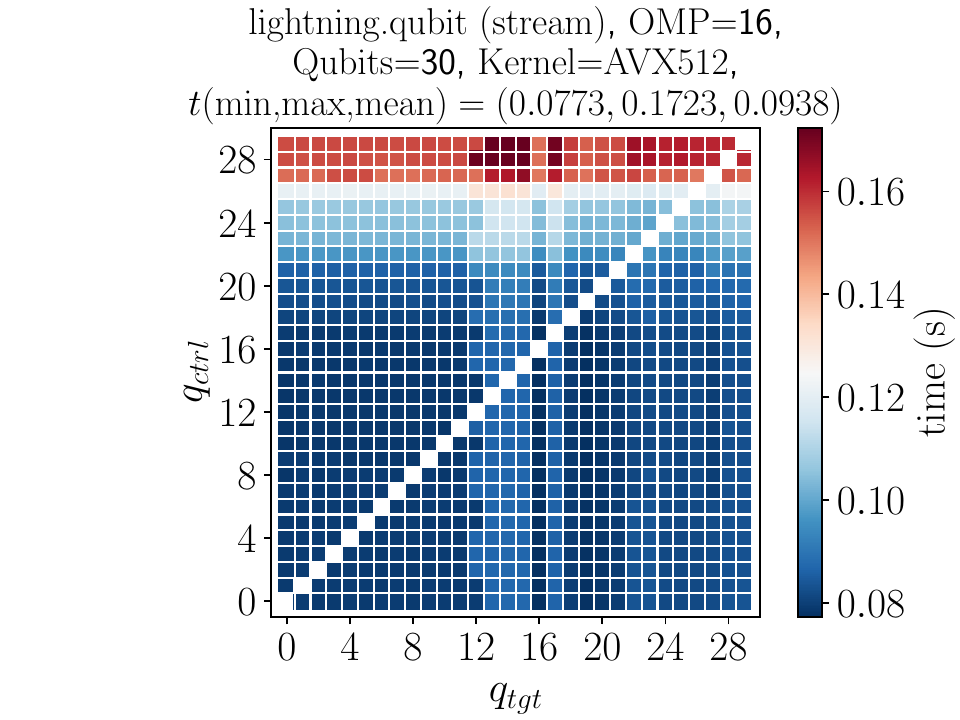} & \includegraphics[width=.24\textwidth, trim={2.5cm 0 0cm 0}, clip]{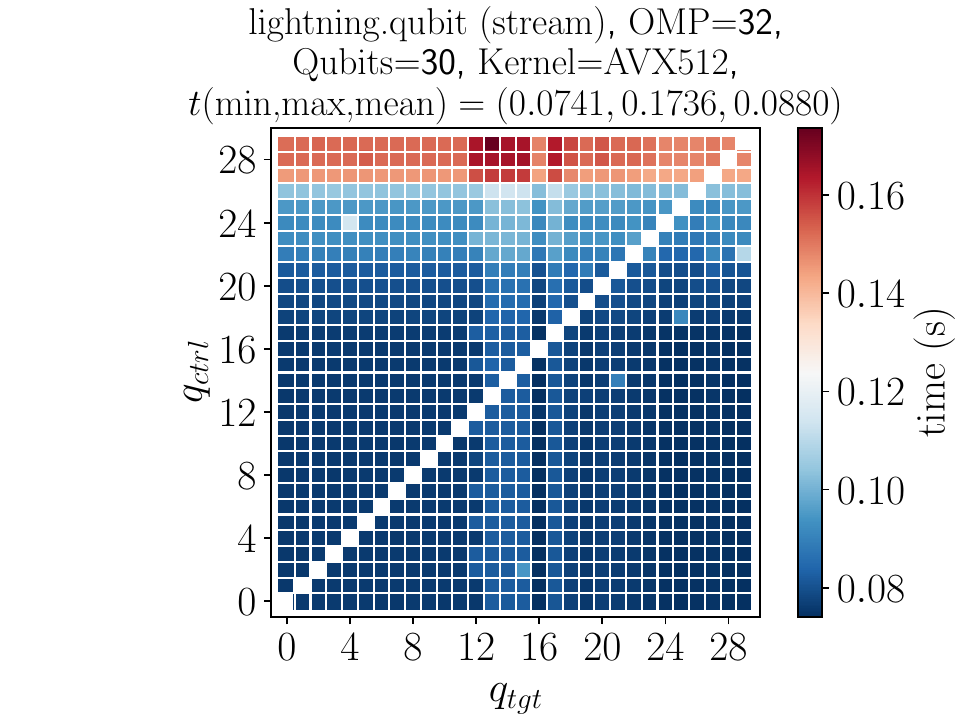}
    \end{tabular}
    \caption{AVX512 (streaming) kernel data, built using the \texttt{bm\_stream} repository branch.}
\end{table}

\clearpage
\subsection{Intel-QS data}

The Intel-QS binaries were built using hash \texttt{f8673c} of the package (July, 2023) against the oneAPI toolkit-provided icpx compiler, with explicitly enabled options for statically compiled binaries, MKL use. Additionally, compiler flags were provided to allow fused multiply-add and AVX-512 auto-vectorization where applicable.

\begin{table}[H]
    \begin{tabular}{c|c|c|c}
        \includegraphics[width=.24\textwidth, trim={2.5cm 0 0cm 0}, clip]{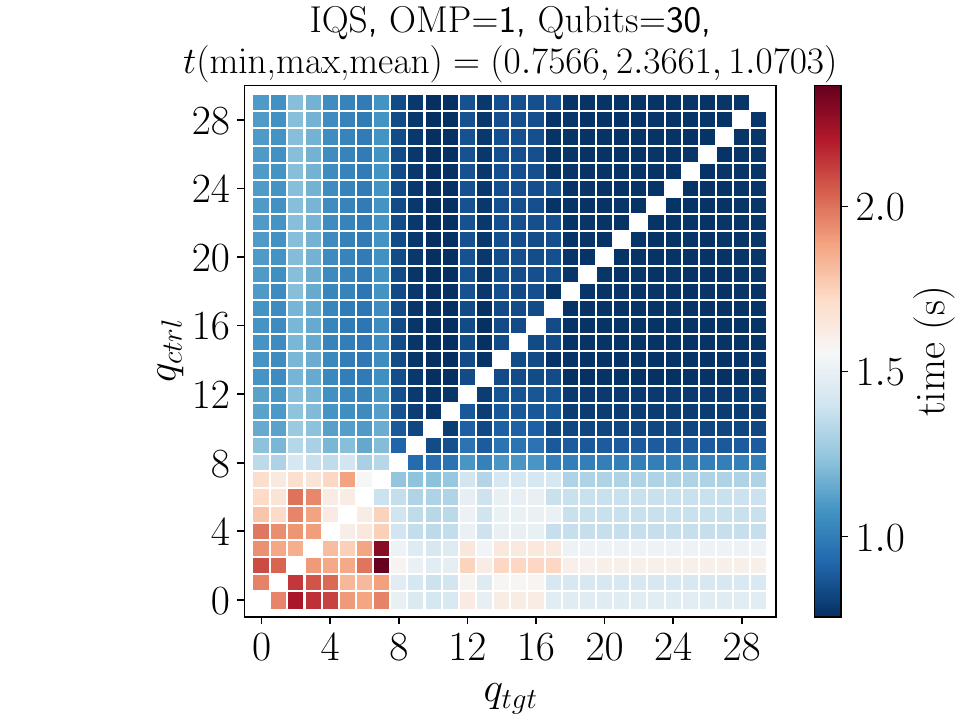} & \includegraphics[width=.24\textwidth, trim={2.5cm 0 0cm 0}, clip]{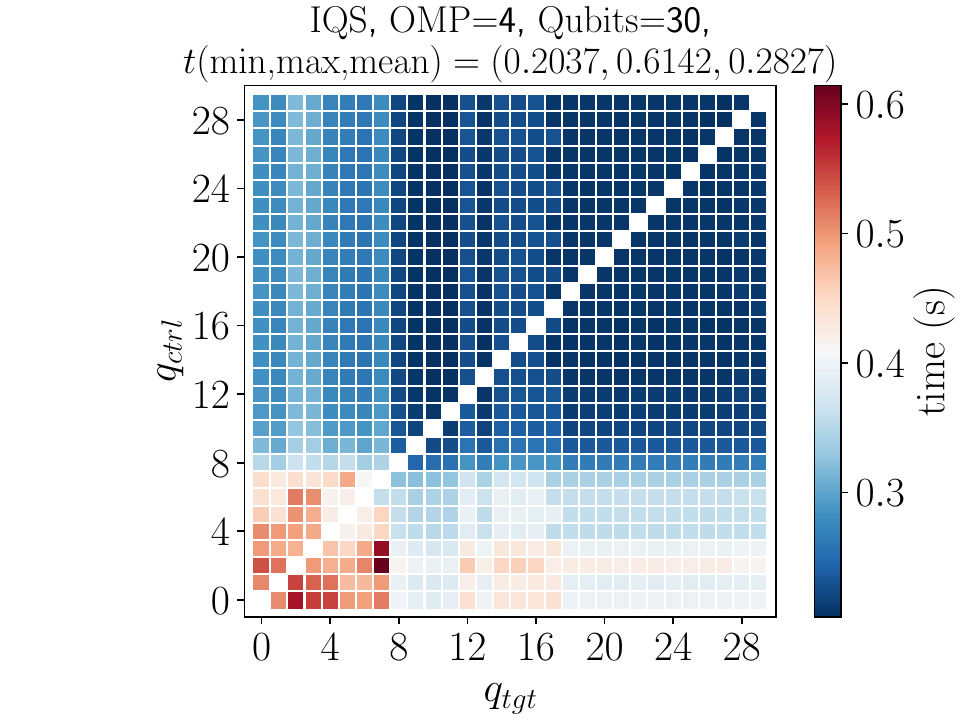} &
        \includegraphics[width=.24\textwidth, trim={2.cm 0 0cm 0}, clip]{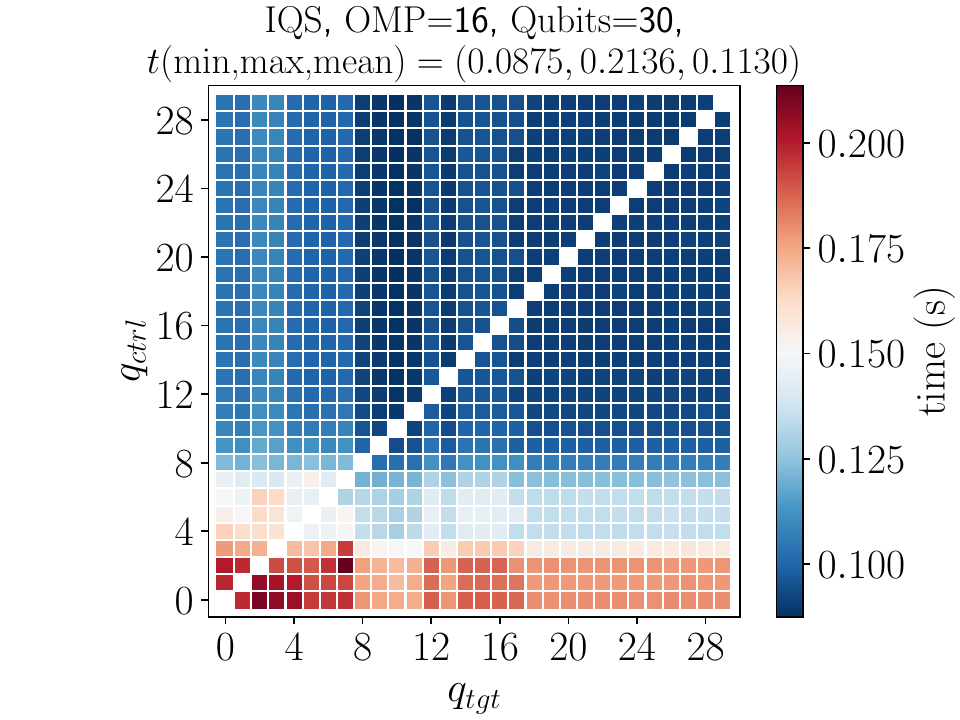} & \includegraphics[width=.24\textwidth, trim={2.3cm 0 0cm 0}, clip]{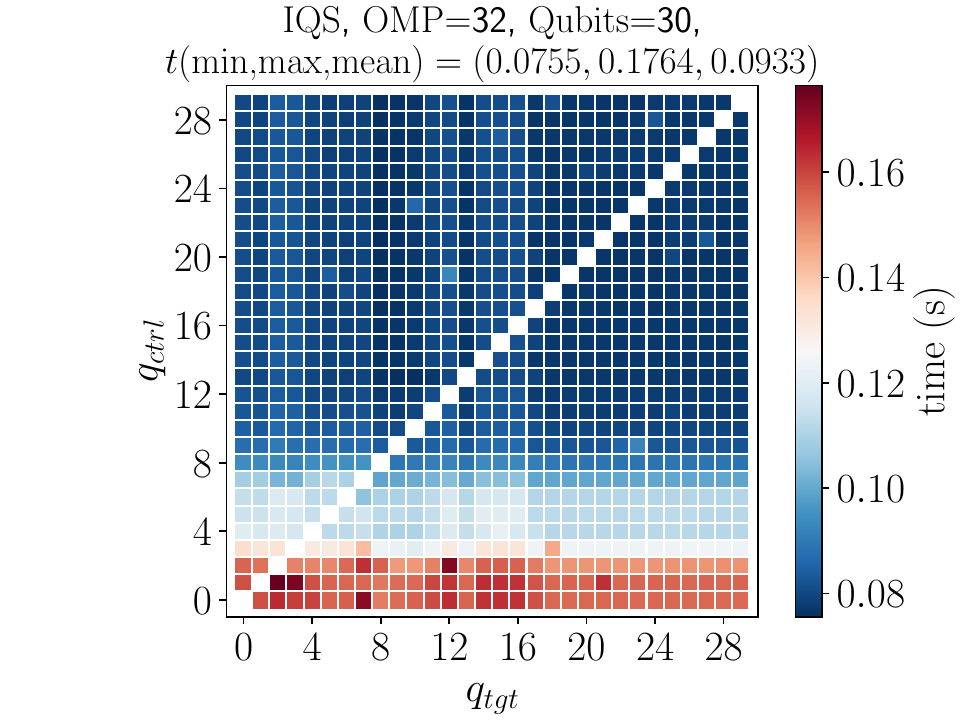}
    \end{tabular}
\end{table}

\subsection{Qulacs data}

The Qulacs binaries were built using version v0.6.2 of the package against GCC 11, with explicitly enabled options for OpenMP and SIMD support. Additionally, compiler flags were provided to allow fused multiply-add and AVX-512 auto-vectorization where applicable.

\begin{table}[H]
    \begin{tabular}{c|c|c|c}
        \includegraphics[width=.24\textwidth, trim={2.5cm 0 0cm 0}, clip]{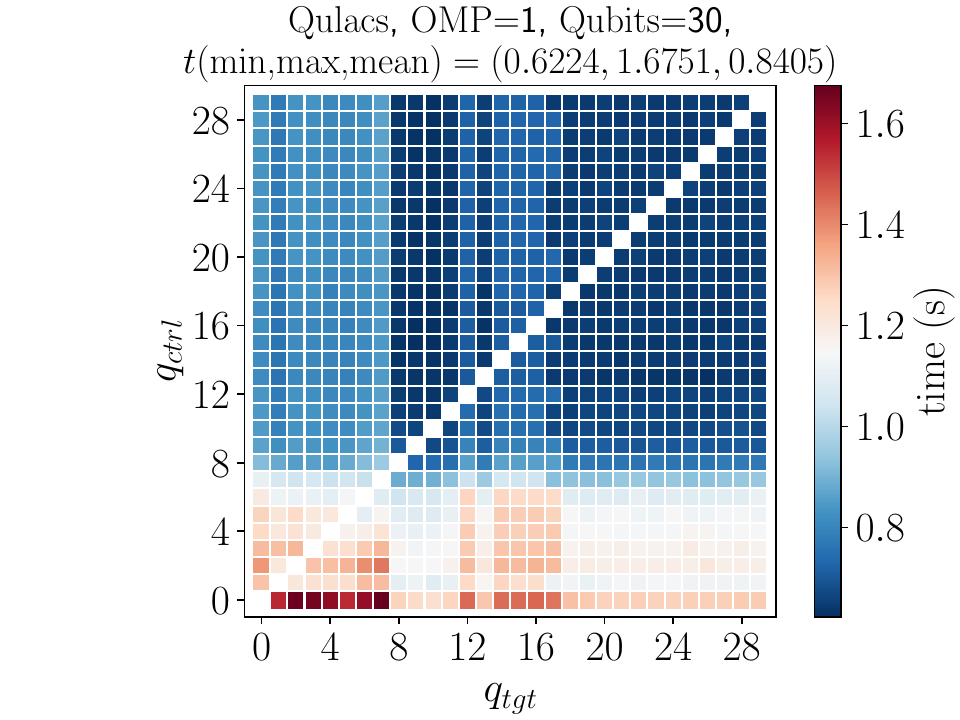} & \includegraphics[width=.24\textwidth, trim={2.3cm 0 0cm 0}, clip]{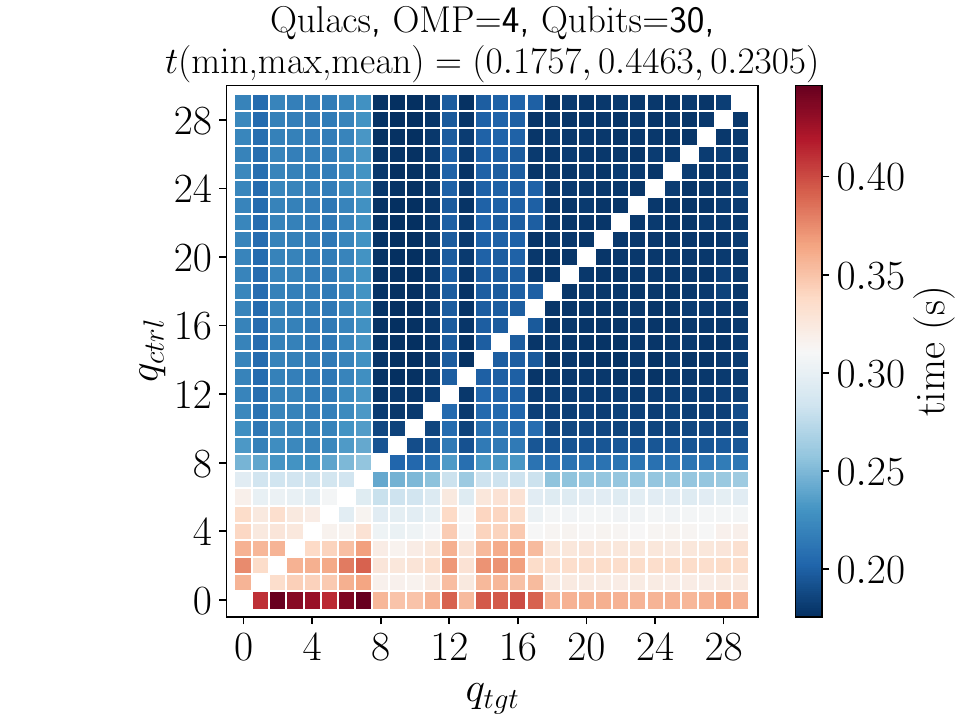} & \includegraphics[width=.24\textwidth, trim={2.3cm 0 0cm 0}, clip]{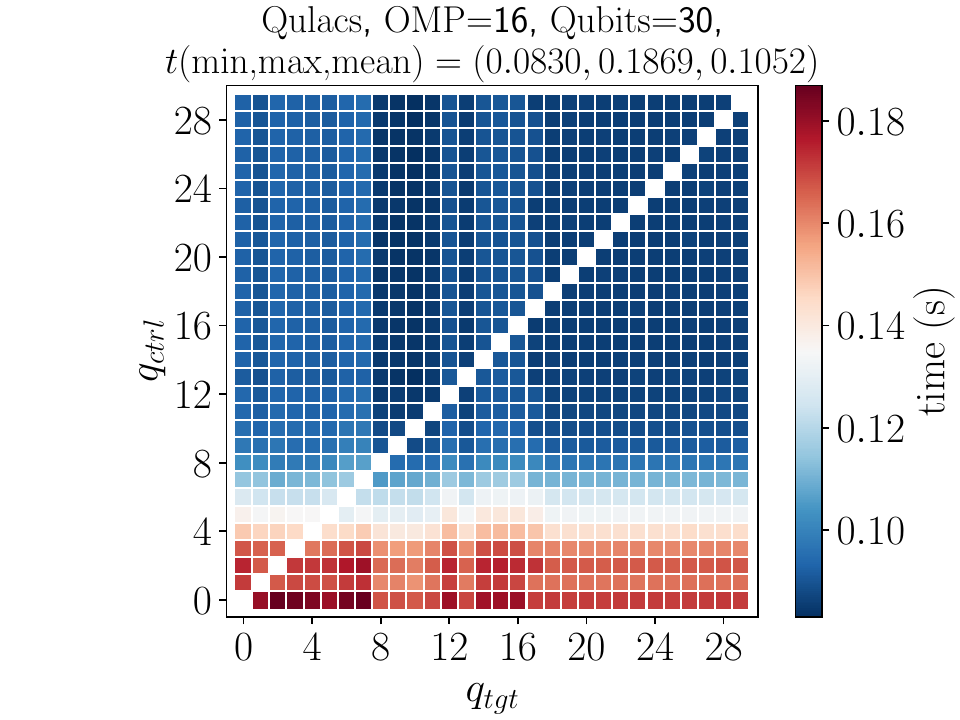} & \includegraphics[width=.24\textwidth, trim={2.3cm 0 0cm 0}, clip]{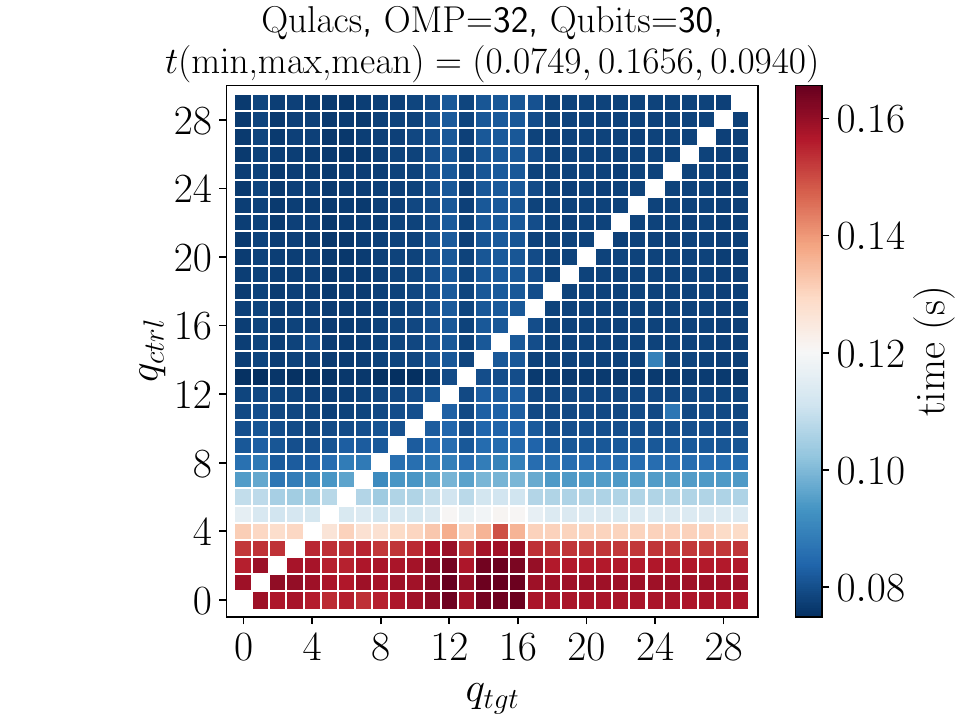}
    \end{tabular}
\end{table}

\subsection{QuEST data}

The QuEST binaries were built using version v3.7.0 of the package against GCC 11. Additionally, compiler flags were provided to allow fused multiply-add and AVX-512 auto-vectorization where applicable.

\begin{table}[H]
    \begin{tabular}{c|c|c|c}
        \includegraphics[width=.24\textwidth, trim={2.5cm 0 0cm 0}, clip]{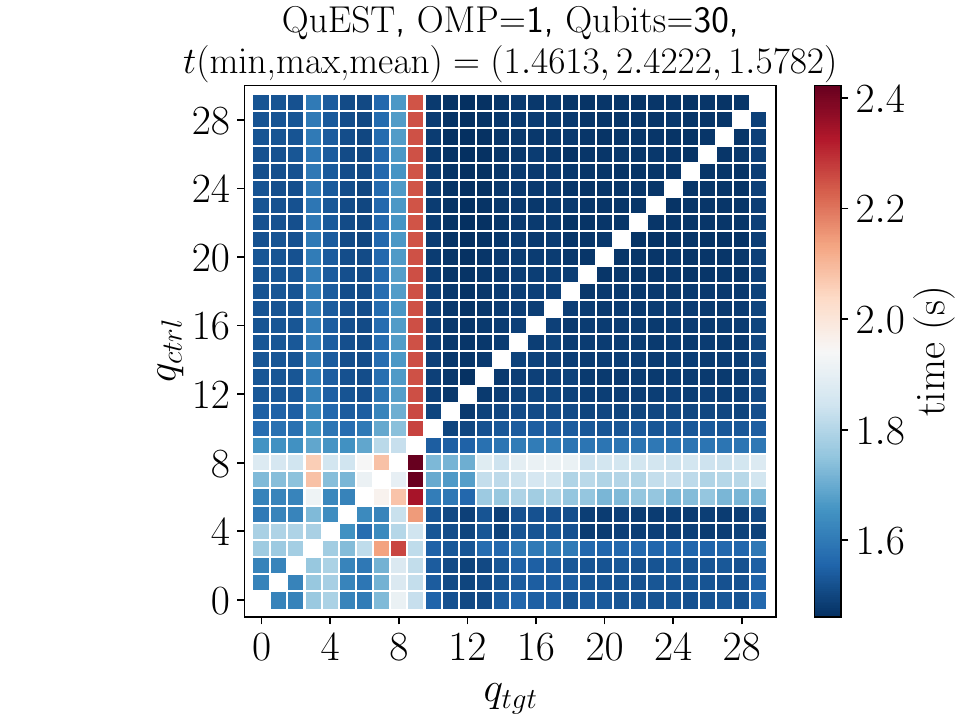} & \includegraphics[width=.24\textwidth, trim={2.5cm 0 0cm 0}, clip]{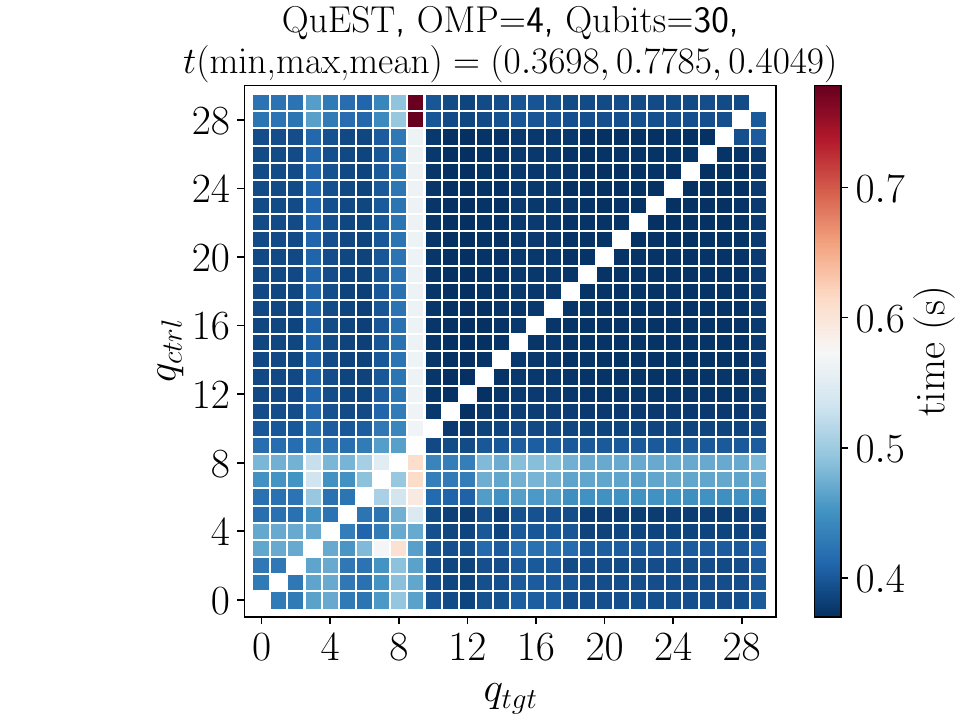} & \includegraphics[width=.24\textwidth, trim={2.3cm 0 0cm 0}, clip]{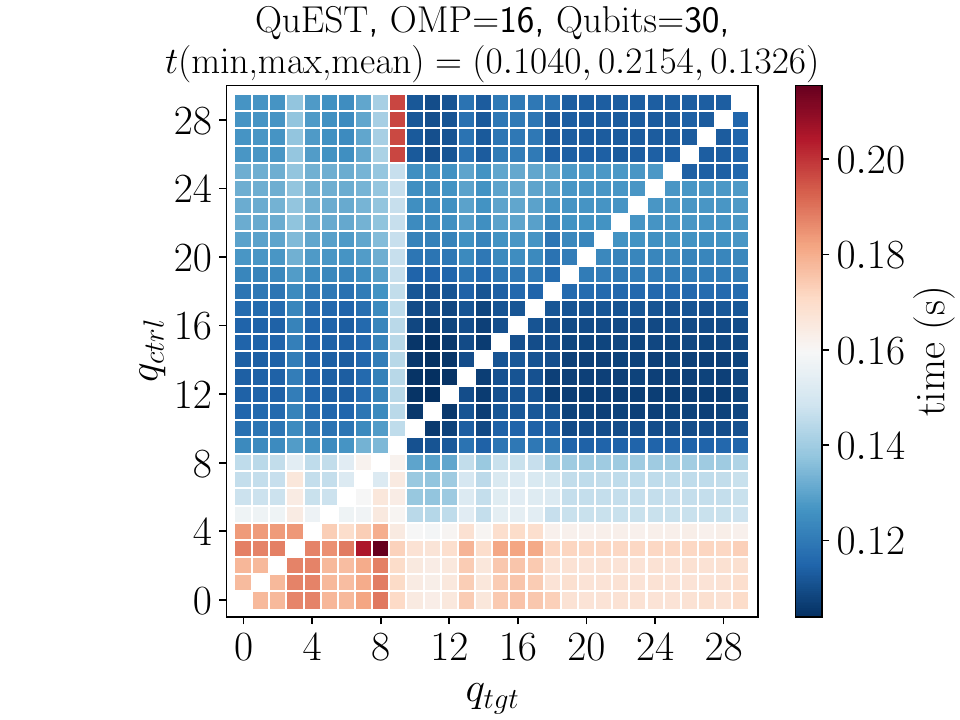} & \includegraphics[width=.24\textwidth, trim={2.3cm 0 0cm 0}, clip]{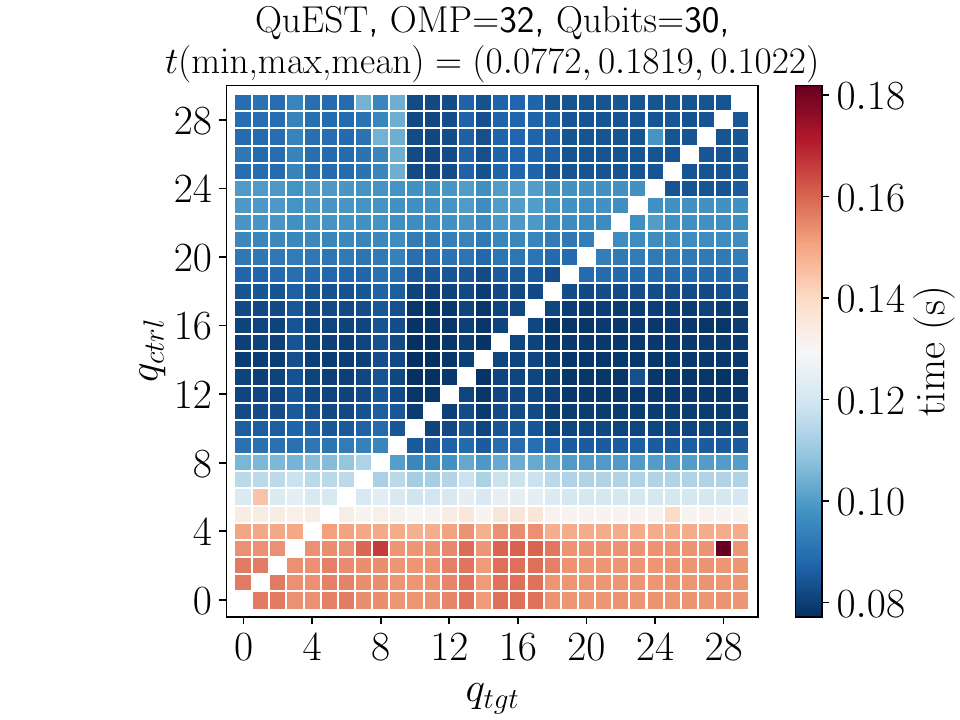}
    \end{tabular}
\end{table}

\end{document}